\documentclass[acmtog,nonacm]{acmart}

\AtBeginDocument{%
  \providecommand\BibTeX{{%
    \normalfont B\kern-0.5em{\scshape i\kern-0.25em b}\kern-0.8em\TeX}}}

\usepackage{wrapfig}
\usepackage{bm}
\usepackage{xcolor}
\usepackage{enumitem}
\usepackage{booktabs}
\usepackage{microtype}
\usepackage{multirow}
\usepackage{amsmath}
\usepackage{stfloats}

\newcommand{\proj}{P_U}
\newcommand{\areafunc}{A}
\newcommand{\signedarea}{\overline{A}}
\newcommand{\pixel}[1]{p^{#1}}
\newcommand{\pixelmax}{R}
\newcommand{\pixeltwo}[1]{q^{#1}}
\newcommand{\grayscale}[1]{\widetilde{g}^{#1}}
\newcommand{\pixelflux}[1]{\widetilde{\Phi}^{#1}}
\newcommand{\riratio}{\eta}
\newcommand{\incomingangle}{\alpha}
\newcommand{\outgoingangle}{\beta}
\newcommand{\incomingdir}{\mathbf{a}}
\newcommand{\incomingorthodir}{\mathbf{a}'}
\newcommand{\outgoingdir}{\mathbf{b}}
\newcommand{\outgoingorthodir}{\mathbf{b}'}
\newcommand{\normal}{\mathbf{n}}
\newcommand{\imageloss}{E_{\text{img}}}
\newcommand{\gradloss}{E_{\text{grad}}}
\newcommand{\boundaryloss}{E_{\text{bdr}}}

\newcommand{\singlearealoss}{f_{\text{area}}}
\newcommand{\epsone}{\epsilon_1}
\newcommand{\epstwo}{\epsilon_2}

\newcommand{\singletirloss}{f_{\text{tir}}}
\newcommand{\barrierloss}{E_{\text{barr}}}
\newcommand{\facevtximg}[2]{\widetilde{\mathbf{v}}_{#1}^{#2}}
\newcommand{\facevtx}[2]{{\mathbf{v}}_{#1}^{#2}}
\newcommand{\imgregionproj}{P_I}
\newcommand{\textif}{\text{if}~}
\newcommand{\textotherwise}{\text{otherwise}}
\newcommand{\pixelarea}{A_p}
\newcommand{\powercell}[2]{\mathcal{P}_{#1}^{#2}}
\newcommand{\targetweightone}{\lambda}
\newcommand{\targetweighttwo}{\gamma}
\newcommand{\centerdistterm}{E_{\text{align}}}
\newcommand{\fluxterm}{E_{\text{flux}}}

\newcommand{\otenergy}{H}
\newcommand{\powercellcenter}{\widetilde{\mathbf{c}}}
\newcommand{\centroidfunc}{\mathbf{C}}
\newcommand{\areatargetflux}{\widetilde{\Phi}_T}
\newcommand{\frontz}{z_{\text{front}}}
\newcommand{\focalz}{z_{\text{focal}}}
\newcommand{\frontbackmap}{\mathcal{H}}
\newcommand{\frontvtx}{\mathbf{v}^{\text{f}}}
\newcommand{\backvtx}{\mathbf{v}^{\text{b}}}
\newcommand{\frontheightfunc}{h}

\newcommand{\xfront}{x^{\text{f}}}
\newcommand{\yfront}{y^{\text{f}}}

\newcommand{\zback}{z^{\text{b}}}
\newcommand{\frontnormal}{\mathbf{n}^{\text{f}}}
\newcommand{\pointsource}{\mathbf{q}}
\newcommand{\fronttriangle}{t^{\text{f}}}
\newcommand{\backtriangle}{t^{\text{b}}}
\newcommand{\solidangle}{S}
\newcommand{\luxcorerender}{{LuxCoreRender}}
\DeclareMathOperator{\mae}{MAE}
\newcommand{\pixelerr}{e}
\newcommand{\RN}[1]{%
  \textup{\uppercase\expandafter{\romannumeral#1}}%
}

\newcommand{\shapeoperator}{W}
\newcommand{\sffmat}{\mathbf{M}}
\newcommand{\meancurv}{H}
\newcommand{\facesmoothloss}{E_{\text{face}}}
\newcommand{\fullsmoothloss}{E_{\text{smooth}}}
\newcommand{\laploss}{E_{\text{lap}}}
\newcommand{\tangentbasisvec}{\mathbf{e}}
\newcommand{\tangentvec}{\mathbf{d}}
\newcommand{\basismat}{\mathbf{B}}
\newcommand{\edgeerrfunc}{h}
\newcommand{\edgesmoothloss}{E_{\text{edge}}}
\newcommand{\sfferrfunc}{\delta}
\newcommand{\smoothweight}{\tau}
\newcommand{\paraheading}[1]{\paragraph{#1}}
\newcommand{\addcolspace}{\hspace{0.3em}}
\newcommand{\addhspace}{\hspace{0.4em}}
\begin{document}

\title{End-to-end Surface Optimization for Light Control}


\author{Yuou Sun}
\affiliation{
  \institution{University of Science and Technology of China}
  \city{Hefei}
  \country{China}}
\email{yosun@mail.ustc.edu.cn}

\author{Bailin Deng}
\affiliation{
  \institution{Cardiff University}
  \city{Cardiff}
  \country{United Kingdom}}
\email{DengB3@cardiff.ac.uk}

\author{Juyong Zhang}
\email{juyong@ustc.edu.cn}
\authornote{Corresponding author (\href{mailto:juyong@ustc.edu.cn}{juyong@ustc.edu.cn}).}
\affiliation{%
	\institution{University of Science and Technology of China}
	\city{Hefei}
	\country{China}
}

%% \renewcommand{\shortauthors}{Trovato and Tobin, et al.}


\begin{abstract}
Designing a freeform surface to reflect or refract light to achieve a target distribution is a challenging inverse problem. In this paper, we propose an end-to-end optimization strategy for an optical surface mesh. Our formulation leverages a novel differentiable rendering model, and is directly driven by the difference between the resulting light distribution and the target distribution. 
We also enforce geometric constraints related to fabrication requirements, to facilitate CNC milling and polishing of the designed surface.
To address the issue of local minima, we formulate a face-based optimal transport problem between the current mesh and the target distribution, which makes effective large changes to the surface shape. The combination of our optimal transport update and rendering-guided optimization produces an optical surface design with a resulting image closely resembling the target, while the geometric constraints in our optimization help to ensure consistency between the rendering model and the final physical results.
The effectiveness of our algorithm is demonstrated on a variety of target images using both simulated rendering and physical prototypes.
\end{abstract}






\begin{teaserfigure}
  \includegraphics[width=\textwidth]{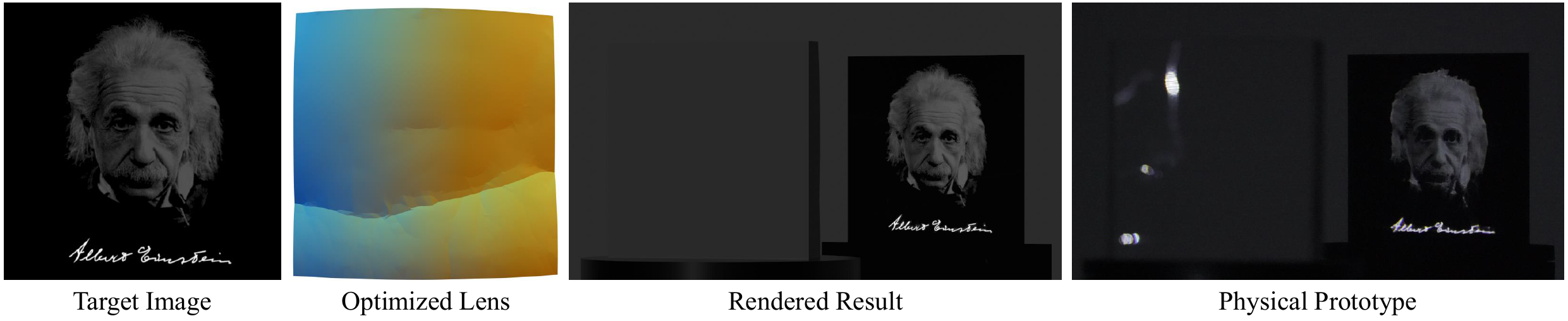}
  \caption{A lens shape designed by our method according to a target image. The light pattern produced by the lens closely resembles the target image, as shown by both the simulated rendering and the physical prototype. Einstein portrait \textcopyright~Magnum/IC photo.}
  \label{fig:teaser}
\end{teaserfigure}

\maketitle

\section{Introduction}

Light plays an important role in our daily life, and people have been learning how to manipulate it for thousands of years. In this paper, we investigate the inverse design of free-form optical surfaces that can exactly transport the light emitted by a source onto a target pattern. It is an important problem in non-imaging optics, and has many applications where one wants to precisely control the direction and density of light rays such as in the field of art, architecture, medical equipment, and energy harvesting.

This problem has been studied in  different fields including computer graphics~\cite{yue2014poisson, schwartzburg2014high, meyron2018light}, computational photography~\cite{kingslake1989history, sun2021end}, and geometrical optics~\cite{budhu2019novel, vsarbort2012spherical}.
In computer graphics, 
existing methods typically first determine certain auxiliary features of the desired optical surface shape, such as its normals~\cite{schwartzburg2014high} or visibility diagram~\cite{meyron2018light}. 
These features are then used to reconstruct the final surface. However, such approaches may fail to produce high-precision final results due to two factors.
Firstly, these methods do not directly consider the difference between the resulting distribution and the target; consequently, errors introduced in feature computation and surface reconstruction may cause notable deviations between the resulting light pattern and the target. Additionally, these methods mainly aim to bring the simulation result close to the target light distribution, without considering the manufacturability of the designed surface.

To overcome these limitations and achieve desirable results in the physically fabricated final surface, we propose a method directly driven by the difference between the resulting pattern and the target pattern, with a rendering model consistent with the light behavior.
Specifically, we adopt a triangle mesh representation for the optical surface, and optimize its shape to produce a reflected/refracted light distribution close to the target.
To this end, we develop a novel differentiable rendering model that considers the reflection/refraction of light on each face according to its face normal, which provides a physically accurate description of the light transport without the need for sampling. Using this model, we formulate an optimization that directly considers the difference between the resulting image and the target image in terms of both pixel values and image gradients. 
In addition, we incorporate the fabrication requirement of the final surface by introducing a piecewise smooth regularization in the optimization, which facilitates the CNC milling and polishing of the final design. Our direct optimization, coupled with an accurate rendering model and geometric constraints, leads to final designs that not only closely match the target images in simulated rendering but also produce good results when physically fabricated.

Due to the nonconvex nature of our optimization formulation, it requires a proper initialization to reach full effectiveness. Therefore, we also develop an optimal transport (OT) based initialization strategy. Although some existing methods \cite{schwartzburg2014high, meyron2018light} have utilized OT to compute the lens shape, they only perform OT once between the source intensity and the target image. Unlike these methods, we formulate an OT problem between the current face-based rendering result and the target image to establish their best correspondence, which is used to update the mesh shape and bring it closer to a desirable solution. A major benefit of our formulation is that the OT can be repeatedly applied during optimization, further improving its effectiveness in steering the mesh shape to avoid local minima.

From the perspective of redistributing light, our face-based OT update and rendering guided optimization are complementary and play different roles. The former establishes a correspondence from each face to a target image region with a matching total flux, producing an approximate face location and orientation to improve the result.
The latter finetunes the mesh shape, in order to faithfully reproduce the details of the target image. 
Therefore, by combining these two techniques in an iterative manner, our final optical surface shape can produce an image with a close resemblance to the target.

We evaluate our method using both digital models and physical prototypes created via CNC milling. Experiments show that our approach can produce optical surfaces with high-fidelity reproduction of the target images, with a notable improvement in precision over existing approaches.
Our work provides an effective solution for high-precision optical surface design for light control, which may benefit other application domains beyond computer graphics.

To summarize, our contributions include:
\begin{itemize}[leftmargin=*]
\item A differentiable rendering model that accurately describes the resulting image for an optical surface represented as a mesh. 
\item An end-to-end algorithm that directly optimizes the optical surface shape, which reduces the difference between the target and result images while enforcing fabrication requirements, leading to a design that can precisely reproduce the target image.
\item A face-based OT initialization strategy, which helps to avoid local minima and enable effective optimization.
\end{itemize}

\section{Related Work}

\paraheading{Inverse surface design for light control}

Optics design aims to shape optical components to create specific visual effects under given lighting conditions.
Lenses, in particular, have been designed for various applications, such as vehicle headlamps \cite{zhu2013optical}, LCD backlights \cite{zhu2019compact}, and projectors \cite{zhao2013integral, damberg2016high}. For example, to achieve varied illumination in industrial settings, \citet{ding2008freeform} devised a lens that transforms LED light sources into rectangular illumination with better uniformity, while the lens developed by \citet{chen2012freeform} can alter it to parallel light.

Apart from lens design for daily usage, another line of work determines the geometry of lens surface that can form a particular caustic image through reflection or refraction.
One of the earliest works is~\cite{finckh2010geometry}, which determines the surface geometry from caustics using the simultaneous perturbation stochastic approximation optimization algorithm that is computationally demanding.
\citet{papas2011goal} employ the EM algorithm instead to optimize micro-surfaces that distribute light onto Gaussian kernels representing the target image. However, their results can suffer from artifacts due to discretization and have difficulties handling low-intensity areas. Other researchers have simplified this problem using different techniques. Some construct the Monge-Ampère equation based on local energy conservation~\cite{wu2013mathematical}, whereas others represent the lens by integrating a normal field~\cite{kiser2013architectural} or formulating it as a Poisson reconstruction problem~\cite{yue2014poisson}. All approaches assume the lens shape as a continuous and smooth surface, significantly reducing the artifacts and improving image quality. 
\citet{schwartzburg2014high} further addressed the problem of generating high-contrast images which cannot be achieved through globally smooth surfaces. 
By utilizing optimal transport, they establish a non-bijective mapping between the incident light field and the target distribution, which is used to compute the surface normals and recover the lens shape. 
Their approach can produce entirely black regions, which was impossible with previous methods.
\citet{Damberg:15} revisited this challenge from the perspective of phase-only spatial light modulation, treating the spatial phase modulator as a freeform lens and developing an efficient optimization algorithm.
\citet{meyron2018light} developed a parameter-free algorithm for caustic design.
They utilize visibility diagrams to find a solution for optimal transport, which balances the distribution of light intensity with energy conservation.
By intersecting a 3D Power diagram with planar or spherical domains, their method can directly obtain the surface shape.

Some other works have focused on reflective properties instead of refractive ones. Among them, \citet{weyrich2009fabricating} achieved target patterns by designing a heightfield mirror. 
\citet{wu2022computational} manipulated the shape and texture of a saucer to directly display an image on its surface, while also reflecting another image on a mirror cup. 
\citet{Shen2023diffGlints} computed a mirror with engraved scratches, which can display distinct images when viewed from different angles. 
\citet{tojo2023stealth} optimized the surface shape to fulfill diverse surface reflectivity requirements.

It is worth mentioning that the concept of inverse design can also be extended to wave optics. \citet{peng2017mix} formed a holographic reconstruction problem based on the Gerchberg-Saxton framework. In addition, \citet{jang2020design} designed the first full-color caustic generator using a freeform holographic optical element.

\begin{figure*}[t]
\centering
\includegraphics[width=\textwidth]{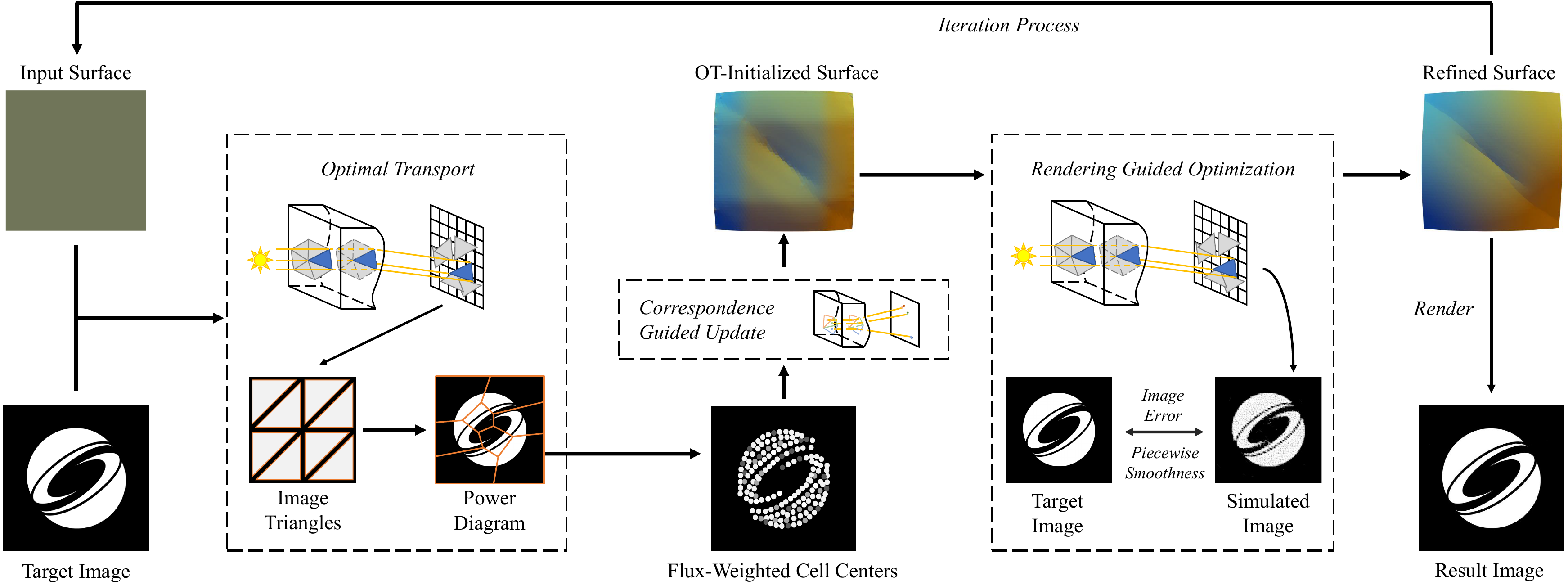}
\caption{Pipeline of our algorithm. We iterate between a face-based optimal transport that adjusts correspondence between mesh faces and target image regions, and a rendering guided optimization that improves the details of the resulting image and reduces its difference from the target.}
\label{fig:pipeline}
\end{figure*}

\paraheading{Differentiable rendering}
Rendering simulates the appearance of a 3D scene based on lighting and object properties. It can be used to determine the light pattern resulting from a lens design.
Unfortunately, traditional rendering procedures contain non-differentiable elements like rasterization and z-buffers, making it challenging to optimize object properties based on the simulated outcome directly. Recently, various research has been devoted to making the rendering process differentiable.
\citet{loper2014opendr} approximated the gradient between pixel-to-vertex coordinates as the image color gradient. However, this method could not propagate gradients into occluded triangles, limiting vertices to receive gradient information solely from nearby pixels.
\citet{liu2019soft} proposed soft rasterization, which leverages probability distributions and aggregation functions to make the rendering pipeline differentiable. 
As a versatile renderer, Mitsuba~2~\cite{NimierDavidVicini2019Mitsuba2} not only takes into account 
 the complex and intricate properties of light but also ensures the rendering process is differentiable and adaptable to diverse applications.
\citet{wang2022differentiable} presented a lens design pipeline with the help of differentiable ray tracing and deep lens systems. The methodologies of both \cite{NimierDavidVicini2019Mitsuba2} and \cite{wang2022differentiable} can be applied to various reverse design problems including caustic design as mentioned above. 

\paraheading{Optimal transport}
The lens design problem can also be treated as a problem of light redistribution. An effective tool for resolving density distribution correspondences is optimal transport (OT), which finds the most efficient way of transferring one mass distribution to another based on a cost function. 
This problem was initially explored by \citet{monge1781memoire}, and later generalized by \citet{kantorovich1942translation} to allow for mass splitting. 
One variant of OT particularly relevant to our problem is the semi-discrete OT, where one distribution is discrete and the other is continuous~\cite{merigot2011multiscale, levy2015numerical}. 
Similar to~\cite{schwartzburg2014high} and \cite{meyron2018light}, we also utilize OT for lens design. However, unlike their continuous OT between the source and target distribution, we use a semi-discrete OT for correspondence between mesh faces and a target image.

\section{Method}
Given a light source and a target image and the refractive index of the lens material, we would like to design a lens shape that refracts the light to form an image on a receptive plane that is close to the target image.
In the following, we first present our algorithm for a parallel light source. Our pipeline is shown in Fig.~\ref{fig:pipeline}. Afterward, we discuss its extension to accommodate a point light source.

\subsection{Measuring light intensity}
\label{sec:lightintensity}

\begin{figure}[b]
\centering
\includegraphics[width=0.98\linewidth]{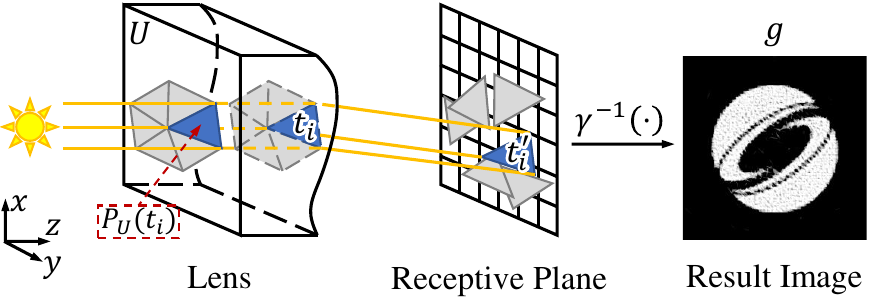} 
\caption{Our rendering process. The light rays pass through the front surface $U$ of the lens without refraction. At each triangle $t_i$ on the back surfaces, the light rays refract according to the normal of $t_i$, forming a corresponding triangle $t'_i$ on the receptive plane. The total flux of $t'_i$ is determined by the area of the projection $\proj(t_i)$ from $t_i$ to $U$. These refracted triangles collectively form the resulting image $g$ after applying the inverse gamma correction $\gamma^{-1}(\cdot)$  (see Eq.~\eqref{eq:image_term}).}
\label{fig:rendering}
\end{figure}

\paraheading{Source flux}
For parallel light, we follow existing methods and assume the front side of the lens (which faces the light source) is flat and perpendicular to the light direction (see Fig.~\ref{fig:rendering}). This avoids refraction by the front surface, thus we only need to  optimize the back surface. We then identify the front surface with a planar domain $U \subset \mathbb{R}^2$, such that the  back surface is a height field over $U$. In this paper, we represent the back surface as a triangle mesh. The intensity of the incident light can be represented as a distribution function $f$ on $U$, where $f(\mathbf{s}) \geq 0 ~ \forall \mathbf{s} \in U$, and  
$
\int_{\mathbf{s} \in U} f(\mathbf{s}) d\mathbf{s} = 1.
$
Since the light does not change direction at the front surface, the flux passing through a triangle  $t_i$ on the back surface can be computed as 
\begin{equation} 
\label{source_flux}
\Phi_i = \int_{\mathbf{s} \in \proj(t_i)} f(\mathbf{s}) d\mathbf{s}, 
\end{equation}
where $\proj(\cdot)$ denotes the projection onto  $U$. For uniform light (i.e., $f(\mathbf{s})$ is constant), $\Phi_i = \frac{\areafunc(\proj(t_i))}{\areafunc(U)}$ where  $\areafunc(\cdot)$ denotes the area.

\paraheading{Target flux}
Given a grayscale target image, we compute a value $\pixelflux{j}$ for each pixel $\pixel{j}$ to describe its flux distribution, such that $\sum_{j=1}^{n_p}\pixelflux{j} = 1$ where $n_p$ is the number of pixels.
Note that the flux of pixel $\pixel{j}$ is represented not by its grayscale value $\grayscale{j}$, but by $\gamma(\grayscale{j})$ where $\gamma(\cdot)$ is the gamma correction. Hence, we compute $\pixelflux{j}$ as
\begin{equation}
    \pixelflux{j}  = \frac{\gamma(\grayscale{j})}{\widetilde{G}},
    \label{target_flux}
\end{equation}
where $\widetilde{G} = \sum_{j =1}^{n_p} \gamma(\widetilde{g}^j)$ is the total flux of all pixels.

\subsection{Differentiable rendering}
\label{sec:diffrendering}
Given the back surface mesh, we propose a differentiable model to evaluate the flux distribution in a designated area on the receptive plane with the same pixel resolution as the target image (see Fig.~\ref{fig:rendering}). 
\begin{wrapfigure}{r}[1em]{0.2\columnwidth}
\vspace*{-0.5em}
\hspace*{-2em}
\centering
\includegraphics[width=0.17\columnwidth]{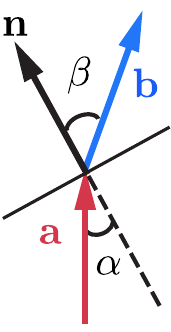}
\label{fig:refraction}
\end{wrapfigure}
At each point of the back surface of the lens, the refracted light ray for an incident ray can 
be determined using Snell's law $\sin \outgoingangle / \sin \incomingangle = {\riratio}$, where $\incomingangle, \outgoingangle$ are the 
angle between the surface normal line and the incident and refracted rays respectively, and  $\riratio$ is the ratio between the refractive indices of the medium on two sides of the interface (see inset figure). The refracted ray direction $\outgoingdir$ can  be computed as (see Appendix~\ref{appx:refraction} for a derivation):
\begin{equation}
    \outgoingdir = \normal \sqrt{1 + \riratio^2((\normal \cdot \incomingdir)^2 - 1)} + \riratio (\incomingdir - (\incomingdir \cdot \normal) \normal),
    \label{eq:RefractedDir}
\end{equation}
where $\incomingdir, \normal$ are unit vectors for the incident ray direction and the surface normal, respectively.
If the back surface is represented as a triangle mesh, then for each face $t_i$ its interior points have the same surface normal
\begin{equation}
\normal_i = \frac{(\facevtx{i}{2}-\facevtx{i}{1}) \times (\facevtx{i}{3} - \facevtx{i}{1})}{\|(\facevtx{i}{2}-\facevtx{i}{1}) \times (\facevtx{i}{3} - \facevtx{i}{1})\|},
\label{eq:FaceNormal}
\end{equation}
where $\facevtx{i}{1}, \facevtx{i}{2}, \facevtx{i}{3}$ are the vertex positions of $t_i$ in the positive orientation.
Since incoming light rays remain parallel after passing through the front surface, it is evident that the outgoing rays from $t_i$ are also parallel. The intersection  between these rays and the receptive plane forms a triangle $t'_i$, where each vertex of $t'_i$ lies on the refracted ray from a corresponding vertex of $t_i$. Assuming that there is no energy loss during refraction and the flux inside $t'_i$ is uniform, we allocate the source flux $\Phi_i$ of $t_i$ to each pixel $\pixel{j}$ on the receptive plane using the intersection area between  $t'_i$ and $\pixel{j}$:
$$
\Phi_i^j= \frac{\Phi_i \cdot {\areafunc(t'_i \cap p^j)}}{{\areafunc(t'_i)}}.
$$
Since both $t'_i$ and $p^j$ are convex polygons, we calculate their intersection $t'_i \cap p^j$ by adopting the convex polygon intersection algorithm from~\cite{o1998computational}, which has linear complexity with respect to the total number of polygon vertices.
Moreover, a non-empty intersection $t'_i \cap p^j$ must also be a convex polygon, and its area $\areafunc(t'_i \cap p^j)$ can be easily calculated in closed form~\cite{o1998computational}.
The total flux $\Phi^j$ for a pixel $\pixel{j}$ is the sum of its allocated flux from all triangles on the back surface:
$$
\Phi^j = \sum_{i = 1}^{n_t} \Phi_i^j,
$$
where $n_t$ is the number of triangles.
For uniform source flux, our model produces a physically accurate flux distribution on the pixels without sampling the light rays.

For each image triangle $t_i'$, we use a scan-line algorithm within its bounding box to determine the pixels it intersects with. 
Moreover, the computation for each triangle is independent and can be carried out in parallel, enabling real-time results.
Our model is differentiable w.r.t. the back surface shape, allowing us to optimize the lens shape using a gradient-based solver, as explained in the next subsection.

Unlike existing differentiable rendering methods \cite{NimierDavidVicini2019Mitsuba2, wang2022differentiable}, our approach utilizes flux-based rendering rather than ray tracing, which allows us to compute the rendering result in closed forms without the need for sampling.

\subsection{Rendering guided optimization}
\label{sec:optimization}
Given a mesh for the back surface, we optimize its vertex positions such that the refracted rays result in an image as close to the target image as possible. This section presents our formulation.

\paraheading{Optimization settings}
We assume the incident light to be in the $+z$ direction. The domain $U$ is chosen as $[0, W] \times [0, H]$ in the $x$-$y$ space for a target image in $W$$\times$$H$ resolution. We triangulate $U$ by sampling uniformly at grid positions and joining them with regular connectivity, and then assign a $z$-coordinate to each vertex (chosen as $0$ by default) to obtain an initial mesh for the back surface. 
The receptive plane is fixed at the plane $z= \focalz$ where $\focalz$ is a user-specified focal length, and its image region is chosen as $[0, W] \times [0, H]$ to align with the back surface. 
The front surface is fixed at the plane $z = \frontz$ where $\frontz$ is specified by the user. (The exact value of $\frontz$ is irrelevant for parallel light but needs to be considered when using a point light source.)
We use the mesh vertex coordinates as optimization variables. To maintain alignment with $U$, we fix the $x$- and/or $y$-coordinates of the boundary vertices at their initial values.

\paraheading{Image terms} 
For closeness between the resulting image and the target, we introduce a term to penalize their pixel value differences:
\begin{equation}
    \imageloss = \sum_{j = 1}^{n_p} (g^j - \widetilde{g}^j)^2,
\label{eq:image_term}
\end{equation}
where $g^j = \gamma^{-1}(\widetilde{G}\Phi^j)$ is the resulting pixel value computed from the flux distribution using Equation (\ref{target_flux}).
Moreover, we use the following term to penalize the difference between their image gradients, in order to preserve the salient features in the target image:
\begin{equation}
\gradloss = \| \mathbf{G}_x - \widetilde{\mathbf{G}}_x \| _F^2 +  \| \mathbf{G}_y - \widetilde{\mathbf{G}}_y \| _F^2,
\end{equation}
where $\mathbf{G}_{x,y}$ and $\widetilde{\mathbf{G}}_{x, y}$ represent the gradient matrices of the resulting image and target images, respectively.
Additionally, we introduce a term to inhibit the refracted rays from reaching a location outside the designated imaging region on the receptive plane:
\begin{equation}
\boundaryloss = \sum_{i = 1}^{n_t} \sum_{k=1}^{3} \| \facevtximg{i}{k} - \imgregionproj(\facevtximg{i}{k})\|^2, 
\end{equation}
where $\facevtximg{i}{k} \in \mathbb{R}^2$ is $k$-th vertex position of the image triangle $t'_i$, and $\imgregionproj(\facevtximg{i}{k}) \in \mathbb{R}^2$ is the closest point to $\facevtximg{i}{k}$ from the imaging region.

\paraheading{Barrier terms}
For the result to be valid, our optimization variables need to satisfy some constraints.
First, as mentioned in Sec.~\ref{sec:lightintensity}, the back surface should be a height field over the domain $U$. Thus the projections $\{\proj(t_i)\}$ of back surface triangles onto $U$ should have no overlap except for the shared edges and vertices. This is equivalent to the condition that the following signed area is positive for each projected triangle:
$$\signedarea\left(\proj(t_i)\right) = \frac{1}{2}\det(\left[ \proj(\facevtx{i}{2}) -  \proj(\facevtx{i}{1}),~ \proj(\facevtx{i}{3}) - \proj(\facevtx{i}{1})\right]).$$
Additionally, very small triangles may cause numerical issues during optimization and should be avoided. Thus we introduce a barrier term $\singlearealoss\left(\signedarea\left(\proj(t_i)\right)\right)$ for the signed area $\signedarea\left(\proj(t_i)\right)$, where
\begin{equation}
    \singlearealoss(a) = 
    \begin{cases}
        0  & \textif{} a \geq \epstwo,\\
        (\frac{\epstwo - \epsone}{a - \epsone}  - 1)^2 & \textif{} \epsone < a < \epstwo,\\
        + \infty & \textotherwise{},
    \end{cases}
    \label{eq:SingleAreaLoss}
\end{equation}
and $\epsone, \epstwo$ are threshold values that satisfy $\epstwo > \epsone > 0$.
Here $\singlearealoss \to +\infty$ when $a \to \epsone$, thus the term ensures each signed area is larger than $\epsone$. Moreover, $\singlearealoss$ decreases to zero when the signed area is larger than  $\epstwo$, to avoid affecting triangles that are large enough. 

In addition,  the refracted ray direction in Eq.~\eqref{eq:RefractedDir} is not well-defined if $1 + \riratio^2((\normal \cdot \incomingdir)^2 - 1) < 0$. In this case, \emph{total internal reflection} occurs and the light is reflected back to the lens interior. We introduce another barrier term $\singletirloss(t_i)$ for each triangle to prevent such scenarios:
\begin{equation}    
    \singletirloss(t_i)
    = 
    \begin{cases}
        \left(1 + \riratio^2((\normal_i \cdot \incomingdir)^2 - 1)\right)^{-1} & \textif{} 1 + \riratio^2((\normal_i \cdot \incomingdir)^2 - 
 1) > 0,\\
        +\infty & \textotherwise{}.
    \end{cases}
\label{eq:tir}
\end{equation}
Interestingly, it can be shown that the term $\singletirloss$ not only prevents total internal reflection but also ensures the refracted light rays do not intersect with the lens surface again (see Apendix~\ref{appx:NoIntersection} for a proof). Thanks to this property, there is no need to consider further refraction, which helps to simplify our formulation.

Applying the two barrier terms to all triangles, we obtain
\[
\barrierloss = \sum_{i=1}^{n_t} \singlearealoss\left(\signedarea\left(\proj(t_i)\right)\right) + \singletirloss(t_i).
\]

\begin{figure}
    \centering
    \includegraphics[width=\columnwidth]{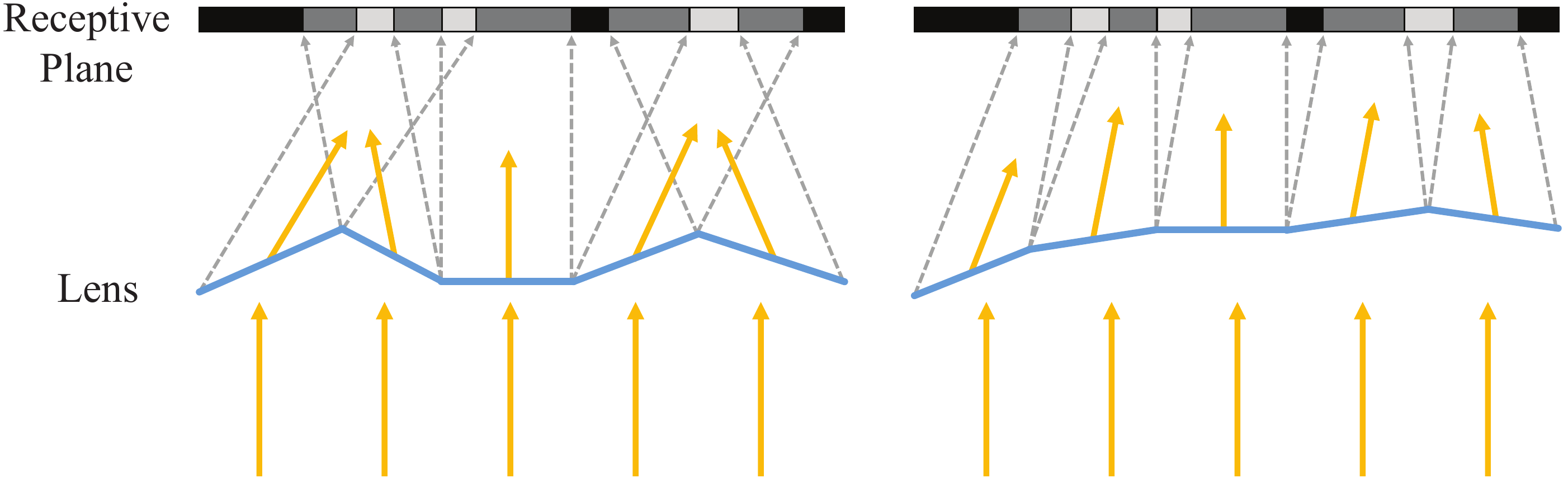}
    \caption{An illustration of different lens shapes that  produce the same distribution of light. We prefer a smoother lens shape as it facilitates fabrication.}
    \label{fig:RoughVsSmooth}
\end{figure}

\paraheading{Smoothness terms}
For a given image, there may be multiple optical surface shapes that can achieve the target distribution of light intensity (see Fig.~\ref{fig:RoughVsSmooth} for a schematic example). We prefer optical surfaces with smoother shapes, since it can be difficult for CNC milling to accurately reproduce a mesh surface with rapid variation of normals within a local area~\cite{zou2021robust}. Additionally, optical surfaces often require polishing as part of their fabrication process. An optical surface design with a rough shape can be more difficult to polish than a smooth surface, especially on a surface represented as a high-resolution mesh: on such surfaces, a rough area may contain small concavities that are not reachable by the polishing tool or small convexities that can be easily affected when the surrounding area is polished; these can lead to under-polishing or over-polishing, both of which will affect the resulting image quality. On the other hand, we cannot expect the surface to be globally smooth in general, as sharp creases may be necessary to achieve some light patterns; for example, if the target image contains two bright areas separated by a pure black stripe, then it cannot be generated from a globally smooth lens surface as the refracted lights from such a lens shape would result in a single continuous bright region. 
Therefore, we need to enforce \emph{piecewise} smoothness for the back surface mesh, such that it is smooth in most areas while allowing for some sharp edges that separate different smooth patches.

A common way to enforce smoothness is to introduce a regularization term for the surface curvature. However, as the mesh vertex positions determine the curvature in a highly non-linear way, directly regularizing the curvature as a function of vertex positions would result in a complex term that is not easy to optimize. Instead, we propose a more tractable approach based on the Weingarten map, a linear operator that captures local surface geometry. Intuitively, the Weingarten map at a point describes how the surface normal changes as we move along different tangent directions. We introduce auxiliary variables on each mesh face to represent the local Weingarten map, and formulate a target function term to adaptively enforce its consistency with the normals of adjacent faces belonging to the same smooth patch; this allows sharp edges to emerge between different patches. Moreover, as the Weingarten map determines the local surface curvature, we further utilize the auxiliary variables to introduce another target function term that regularizes the curvature on each mesh face. These two terms combined enforce the piecewise smoothness of the mesh surface.

More specifically, for a point $\mathbf{p}$ on a smooth surface, the Weingarten map is a linear map $\shapeoperator_{\mathbf{p}}: T_{\mathbf{p}} \mapsto T_{\mathbf{p}}$ within the tangent space $T_{\mathbf{p}}$ at $\mathbf{p}$, and it maps each tangent vector $\tangentvec \in T_{\mathbf{p}}$ to the differential of the surface normal at ${\mathbf{p}}$ along $\tangentvec$.
The map $\shapeoperator_{\mathbf{p}}$ can be represented as a $2 \times 2$ symmetric matrix with respect to an orthonormal basis of $T_{\mathbf{p}}$~\cite{Spivak1999}. Thus, to discretize the Weingarten map, we first choose for each mesh face $t_i$ two orthonormal basis vectors  $\tangentbasisvec_{i,1}$ and $\tangentbasisvec_{i,2}  = \normal_i \times \tangentbasisvec_{i,1}$, where $\tangentbasisvec_{i,1}$ is the unit vector for an edge of $t_i$ and $\mathbf{n}_i$ is the unit face normal as defined in Eq.~\eqref{eq:FaceNormal}. 
Then we introduce three auxiliary variables $a_i, b_i, c_i \in \mathbb{R}$ to represent the symmetric matrix for the Weingarten map under this basis:
\begin{equation}
    \sffmat_i
    = \begin{bmatrix}
        a_i & c_i\\
        c_i & b_i 
    \end{bmatrix}.
    \label{eq:SFFMatrix}
\end{equation}
\begin{wrapfigure}{r}[0.5em]{0.5\columnwidth}
\vspace*{-0.9em}
\hspace*{-3em}
\centering
\includegraphics[width=0.49\columnwidth]{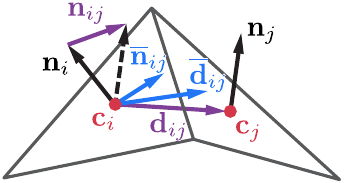}
\vspace*{-0.9em}
\end{wrapfigure}
The matrix $\sffmat_i$ needs to be consistent with the face normals in the neighborhood of $t_i$. We discretize this condition in a way similar to~\cite{rusinkiewicz2004estimating}.
First, we associate the normal $\mathbf{n}_i$ of each face $t_i$ to its centroid $\mathbf{c}_i$.  Then, for each adjacent face $t_j$ that shares a common edge with $t_i$, we compute the difference vectors
\begin{equation}
    \mathbf{d}_{ij} = \mathbf{c}_j - \mathbf{c}_i, \quad 
    \mathbf{n}_{ij} = \mathbf{n}_j - \mathbf{n}_i,
    \label{eq:DiffVectors}
\end{equation}
between the positions and normals of their centroids, and project them onto the tangent plane of $t_i$ to derive their local coordinates
\begin{equation}
    \overline{\mathbf{d}}_{ij} = \basismat_i^T \mathbf{d}_{ij},
    \quad
    \overline{\mathbf{n}}_{ij} = \basismat_i^T \mathbf{n}_{ij},
    \label{eq:ProjDiffVectors}
\end{equation}
where $\basismat_i = [\tangentbasisvec_{i,1}, \tangentbasisvec_{i,2}] \in \mathbb{R}^{3 \times 2}$ is the tangent basis matrix for $t_i$ (see the inset figure above). The matrix $\sffmat_i$ should map $\overline{\mathbf{d}}_{ij}$ to $\overline{\mathbf{n}}_{ij}$ approximately.
Thus, we measure the consistency between $\sffmat_i$ and the geometry of $t_i$ and $t_j$ via the term
\begin{equation}
    \sfferrfunc(t_i, t_j) =  \frac{\|\sffmat_i \overline{\mathbf{d}}_{ij} -  \overline{\mathbf{n}}_{ij}\|^2}{\|\overline{\mathbf{d}}_{ij}\|^2}.
    \label{eq:ConsistencyTerm}
\end{equation}
In addition, if $t_i$ and $t_j$ belong to the same smooth patch, then the geometry of $t_i$ and $t_j$ should be consistent with the Weingarten matrix at $t_j$ as well. Thus, we define the consistency error term for the common edge $e_{ij}$ between $t_i$ and $t_j$ as
\[
    \edgeerrfunc(e_{ij}) =  \sfferrfunc(t_i, t_j) + \sfferrfunc(t_j, t_i).
\]
To achieve piecewise smoothness, we should reduce the value of $\edgeerrfunc$ on most edges, while allowing for large values on some sharp edges that separate different smooth patches. In other words, we need to achieve sparsity of the values $\{\edgeerrfunc(e_{ij})\}$. To this end, we use the Welsch function~\cite{holland1977robust} as a robust norm to promote sparsity, and derive an edge smoothness term as
\[
  \edgesmoothloss = \sum_{e_{ij} \in \mathcal{E}_I} \Psi_\nu(\sqrt{\edgeerrfunc(e_{ij})}),
\]
where $\mathcal{E}_I$ denotes the set of internal mesh edges, $\Psi_\nu$ is the Welsch function 
\begin{equation}
    \Psi_\nu(x) = 1 - \exp(-\frac{x^2}{2 \nu^2}),
    \label{eq:WelschFunc}
\end{equation}
and $\nu$ is a user-specified parameter. The Welsch function attains the minimum value at $x = 0$, is monotonically increasing on $[0, +\infty)$, and is bounded from above by $1$. In this way, it penalizes the errors $\{\edgeerrfunc(e_{ij})\}$ while limiting the influence of large error values, thus allowing for sharp edges.

Moreover, on a smooth surface, the mean curvature at a point equals half the trace of its Weingarten map~\cite{doCarmo1976differential}. Thus we derive a discrete mean curvature measure for the face $t_i$ from the matrix $\sffmat_i$ as
\[
    \meancurv_i = \frac{a_i + b_i}{2}.
\]
Using this relation, we define a surface smoothness term that penalizes the mean curvature:
\begin{equation}
    \facesmoothloss = \sum_{i=1}^{n_t} (\meancurv_i)^2 A_i,
    \label{eq:FaceSmoothness}
\end{equation}
where $A_i$ is the area of face $t_i$. Note that $\facesmoothloss$ can be considered as a discrete integral of the squared mean curvature over the mesh.
We combine $\facesmoothloss$ and the term $\edgesmoothloss$ introduced previously to enforce smoothness within each patch while allowing for sharp edges to emerge between adjacent patches.

Finally, to avoid poor triangulation, we also introduce a Laplacian smoothness term for the projections of the mesh vertices $\{\mathbf{v}_j\}$ onto the height field domain $U$:
\begin{equation}
    \laploss = \sum_{j \in \mathcal{V}_{I}} \left\|\proj(\mathbf{v}_j) - \frac{1}{|\mathcal{N}_j|}\sum\nolimits_{i \in \mathcal{N}_j} \proj(\mathbf{v}_i)\right\|^2,
\end{equation}
where $\mathcal{V}_{I}$ denotes the index set of interior vertices, and $\mathcal{N}_j$ is the index set of neighbor vertices for vertex $j$.

Our overall smoothness term is a weighted sum
\begin{equation}
    \fullsmoothloss = \facesmoothloss + \smoothweight_1 \edgesmoothloss + \smoothweight_2 \laploss.
    \label{eq:FullSmoothnessLoss}
\end{equation}

\paraheading{Overall formulation and numerical solving }
Combining the terms proposed above, we derive an optimization problem
\begin{equation}
    \min~~ \targetweightone_1 \imageloss  + \targetweightone_2 \gradloss + \targetweightone_3 \boundaryloss + \targetweightone_4 \fullsmoothloss +
    \targetweightone_5 \barrierloss,
    \label{eq:TriangleOptimization}
\end{equation}
where $\{\targetweightone_i\}$ are user-specified weights. 
We solve it with an L-BFGS solver~\cite{liu1989limited}. For efficiency, we evaluate the term values and gradients in parallel on the GPU, and use automatic differentiation to compute the derivatives. 
Additionally, the starting mesh must satisfy the conditions enforced by the barrier terms. This is achieved with a flat surface parallel to the front.

Some existing approaches~\cite{yue2014poisson, schwartzburg2014high, meyron2018light} indirectly optimize the lens surface mesh using auxiliary features (such as normal fields) derived from the target image, which can lead to accumulated errors. In contrast, our optimization is directly driven by the difference between the rendered image and the target, both in terms of per-pixel intensity and image gradient. This end-to-end formulation allows fine-grained optimization of the surface to accurately reproduce the target.

\subsection{Face-based optimal transport initialization}
\label{sec:initialization}
Due to the nonconvex nature of our optimization problem, a proper initialization is often necessary to help obtain an optimal solution. This can be seen in the first column of Figure \ref{fig:ablation_study}, where initializing the mesh with a trivial flat surface produces unsatisfactory results.
This is because L-BFGS is inherently a local solver that makes a relatively small change in each iteration, and may stop at a local minimum before reaching a better solution far away from the initial point.
Therefore, we need a procedure to make larger changes and move the current solution closer to a desirable result, to avoid getting stuck at local minima. 
Similar to~\cite{schwartzburg2014high}, we consider our lens shape optimization as re-allocating the flux on the receptive plane to achieve a target distribution,  and use optimal transport (OT) to determine a correspondence that guides the allocation. However, unlike~\cite{schwartzburg2014high} that perform OT between the source and target flux distributions, we formulate an OT problem between the current face-based distribution and the target, and modify the current mesh accordingly.

\begin{figure}[t!]
	\centering
	\includegraphics[width=1.0\columnwidth]{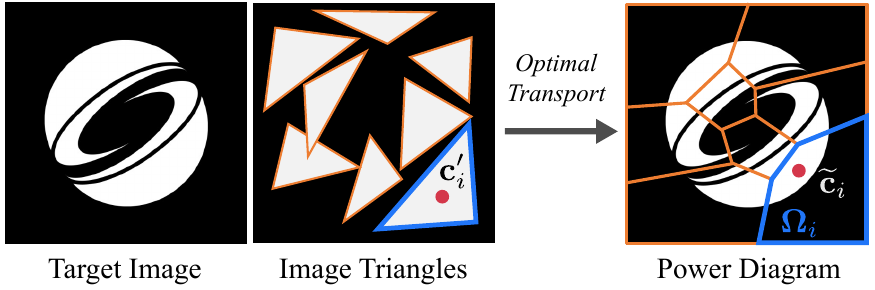}
	\caption{An illustration of our discrete optimal transport process. 
	Given the current image triangles resulting from our rendering model, we associate the total flux of each triangle to its centroid $\mathbf{c}'_i$, and optimize a partition of the target image such that the flux at $\mathbf{c}'_i$ can be minimally transported to its corresponding cell $\Omega_i$ in the partition to match the target image.
	Then we compute the flux-weighted weighted centroid $\widetilde{\mathbf{c}}_i$ of each cell to guide the update of the lens shape.}
	\label{fig:ot}
\end{figure}

\paraheading{Optimal transport}
Recall that each face $t_i$ forms a triangle $t'_i$ on the image plane with a total flux  $\Phi_i$. To facilitate our formulation, we simplify each $t'_i$ to a point at its centroid $\mathbf{c}'_i$ with a flux  $\Phi_i$.
Each pixel $p^j$ of the target image has a total flux $\widetilde{\Phi}^j$ given by Eq.~\eqref{target_flux}, and we consider each point $\mathbf{s}$ in the interior of $p^j$ to have a flux $\widetilde{\Phi}_{\mathbf{s}} = \widetilde{\Phi}^j / \pixelarea$, where $\pixelarea$ is the pixel area.
Consider a partition of the image region $I$ into $n_t$ cells $\{\Omega_i\}$, each corresponding to one point $\mathbf{c}'_i$, so that the flux at $\mathbf{c}'_i$ can be transported to $\Omega_i$ to match the target flux distribution (see Fig.~\ref{fig:ot}), i.e., $\Phi_i$ is the same as the total target flux $\areatargetflux(\Omega_i)$ within $\Omega_i$, where the latter is computed as
\begin{equation}
\areatargetflux(\Omega_i) = \sum_{j = 1}^{n_p} \frac{\widetilde{\Phi}^j \cdot {\areafunc (p^j \cap \Omega_i)}}{\areafunc(p^j)}. 
    \label{eq:target_flux}
\end{equation}
Among all such partitions, we search for one with the least cost of flux transport, which can be described as
\[
\begin{aligned}
\min_{\{\Omega_i\}} & &\sum_{i = 1}^{n_t} \int_{\mathbf{s} \in \Omega_i} \left\| \mathbf{s} - \mathbf{c}'_i \right\|^2 \widetilde{\Phi}_{\mathbf{s}} d\mathbf{s},\\
    \text{s.t.} & &\Phi_i = \areatargetflux(\Omega_i),~~
    \bigcup_{i = 1}^{n_t} \Omega_i = I.
\end{aligned}
\]
Here the transport cost from  $\mathbf{c}'_i$ to $\Omega_i$ is measured by the integrated squared distance between $\mathbf{c}'_i$ and each point in $\Omega_i$, weighted by the flux at the point.
It has been shown in~\cite{levy2015numerical} that the solution can be represented using a power diagram generated from the points $\mathbf{c}'_1,\ldots,\mathbf{c}'_{n_t}$ with appropriate weights $\mathbf{w} = (w_1, \ldots, w_{n_t})$, such that $\Omega_i = I \cap \powercell{i}{\mathbf{w}}$ where $\powercell{i}{\mathbf{w}}$ is the power cell corresponding to $\mathbf{c}'_i$:
\[
  \powercell{i}{\mathbf{w}}  =\{\mathbf{x} \mid \| \mathbf{x} - \mathbf{c}'_i \|^2 - w_i \leq \| \mathbf{x} - \mathbf{c}'_j \|^2 - w_j, \forall j = 1, \cdots, n_t\}.
\]
The weights $\mathbf{w}$ is the minimizer of a convex energy~\cite{levy2015numerical}:
\[
    \otenergy(\mathbf{w}) = \sum_{i = 1}^{n_t} \left( - \int_{\mathbf{s} \in \powercell{i}{\mathbf{w}}} (\| \mathbf{x} - \mathbf{c}'_i \|^2 - w_i) \widetilde{\Phi}_\mathbf{s} d \mathbf{s} - \Phi_i w_i \right),
\]
whose gradient has the form ${\partial \otenergy(\mathbf{w})}/{\partial w_i} = \areatargetflux(\powercell{i}{\mathbf{w}}) - \Phi_i$.
We minimize $\otenergy(\mathbf{w})$ with an L-BFGS solver, using CGAL~\cite{cgal:eb-23a} to generate the power diagram.
To evaluate $\otenergy$ accurately, we convert it to a quartic polynomial using Green's theorem and Newton-Leibniz formula (see Appendix~\ref{appx:H}).

\begin{wrapfigure}{r}[1em]{0.43\columnwidth}
\vspace*{-1.2em}
\hspace*{-2.5em}
\centering
\includegraphics[width=0.4\columnwidth]{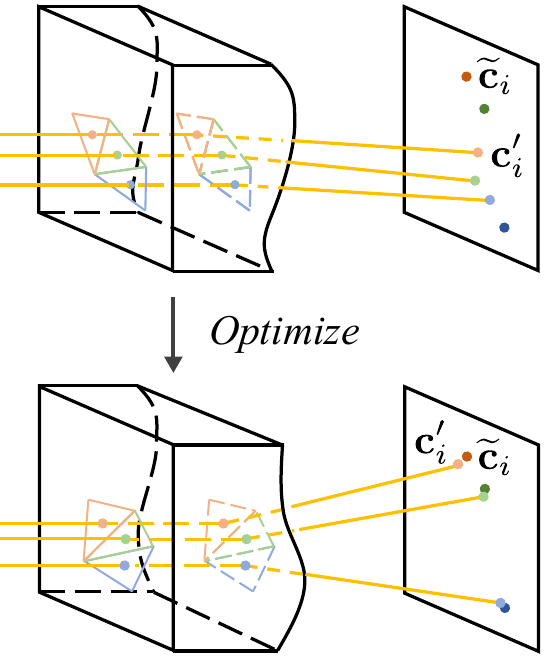}
\end{wrapfigure}
\paraheading{Correspondence guided update}
After deriving the partition $\{\Omega_i\}$ using OT, we update the back surface mesh such that the image $t'_i$ of each triangle aligns better with its corresponding region $\Omega_i$ in terms of both location and total flux, which helps to bring the resulting flux distribution closer to the target (see inset figure). 
To this end, we formulate an optimization 
\begin{equation}
    \min~\targetweighttwo_1 \centerdistterm + \targetweighttwo_2 \fluxterm
    + \targetweightone_3 \fullsmoothloss
    + \targetweighttwo_4 \barrierloss,
    \label{eq:MeshUpdateProblem}
\end{equation}
where  $\{\targetweighttwo_i\}$ are user-specified weights.
Here 
\begin{equation}
    \centerdistterm = \sum_{i=1}^{n_t} \|\mathbf{c}'_i - \powercellcenter_i\|^2
\end{equation} 
enforces alignment between the centroid $\mathbf{c}'_i$ of $t'_i$ and the flux-weighted centroid $\powercellcenter_i$ of $\Omega_i$:
\begin{equation}
\powercellcenter_i = \frac{\sum_{j=1}^{n_p} \centroidfunc(\Omega_i \cap \pixel{j} )~\widetilde{\Phi}^j~\areafunc(\Omega_i \cap \pixel{j} )}{\sum_{j=1}^{n_p} \widetilde{\Phi}^j~\areafunc(\Omega_i \cap \pixel{j} )},
\label{eq:flux_weighted_centroid}
\end{equation}
where $\centroidfunc(\cdot)$ denotes the centroid.
The term 
\begin{equation}
\fluxterm = \sum_{i=1}^{n_t} ( \Phi_i - \Phi_i^{\text{old}} )^2
\end{equation}
maintains the total flux of each triangle, where $\Phi_i^{\text{old}}$ denotes the flux of $t_i$ before the update. 
$\fullsmoothloss$ and $\barrierloss$ are the same as in Eq.~\eqref{eq:TriangleOptimization}. The problem~\eqref{eq:MeshUpdateProblem} is solved using L-BFGS.

\begin{figure*}[t]
	\centerline{\includegraphics[width=\textwidth]{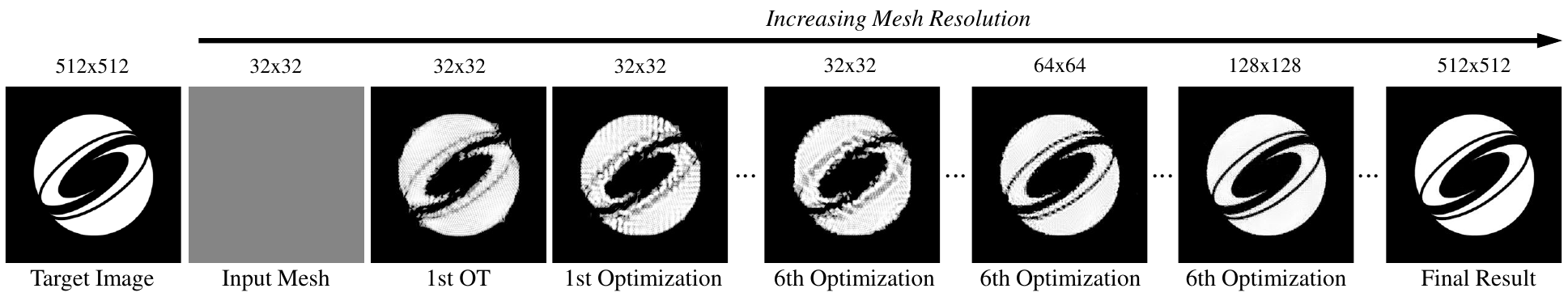}}
	\caption{Our iterative optimization process. We show the resulting images derived from the intermediate results of our optimization process. The numbers on top are the resolution of the target image (the leftmost column) or the intermediate mesh (the other columns).  The label at the bottom indicates the completed numbers of iterations for OT or rendering guided optimization at the current resolution.}
	\label{fig:iteration}
\end{figure*}

Unlike~\cite{schwartzburg2014high} where OT is performed only once, our face-based OT formulation can be repeatedly applied to the mesh. In our implementation, we adopt a coarse-to-fine strategy, starting from a coarse input mesh and a downsampled target image, and then alternate between OT and our rendering guided optimization for six iterations. Then we subdivide the mesh and increase the target image resolution, and repeat the alternation between OT and optimization. This process is repeated until the final resolution is reached. 
In this way, our method first generates a coarse mesh with a resulting image that roughly aligns with the target, then gradually refines and improves its details for closer resemblance as the resolution increases.
Fig.~\ref{fig:iteration} shows an example of this process.  
In addition, following existing work that utilizes the Welsch function as robust norm~\cite{Zhang2022Fast}, we gradually decrease the parameter $\nu$ in Eq.~\eqref{eq:WelschFunc}, to incrementally reduce the influence of large errors and allow sharp edges to emerge during the optimization.

\begin{figure}[t]
\centering
\includegraphics[width=1\linewidth]{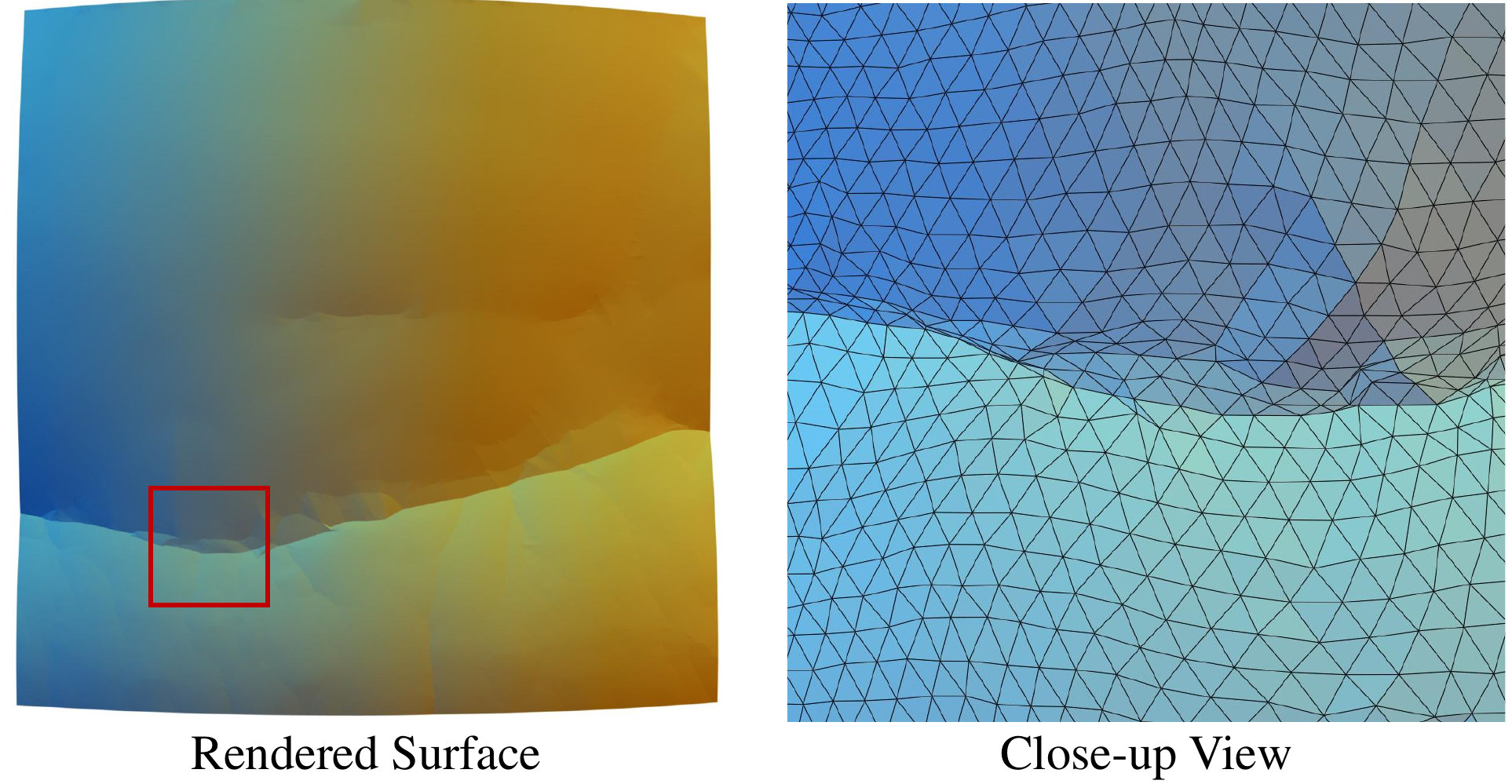}
\caption{The optimized lens shape for the target image in Fig.~\ref{fig:teaser}. The left image shows the full optimized surface, while the right image provides a close-up view of the triangulation in the region highlighted by the red rectangle. The mesh contains sharp edges that are necessary for reproducing the discontinuities in the target image, while maintaining the smoothness within each patch and avoiding poor triangulation.}
\label{fig:close_up}
\end{figure}

Fig.~\ref{fig:close_up} showcases an example of the final optimized lens surface for the target image in Fig.~\ref{fig:teaser}, along with a close-up view of its triangulation. 
Thanks to the smoothness terms and barrier terms in our rendering-guided optimization, our method produces a high-quality mesh representation that contains the sharp edges needed for reproducing the discontinuities in the target image, while maintaining the smoothness within each patch and avoiding poor triangulation.

\subsection{Point light source}
\label{sec:PointLightSource}
Besides refraction of parallel light, our approach can be extended to handle a point light source and reflection. In the following, we present the extension to a point light source. Further discussion about reflection can be found in Appendix~\ref{appx:reflection}.

With a given point light source, the light refracts at the front surface, and we cannot directly obtain the incident light direction for a back surface mesh vertex in a closed form in general: the computation of such a direction would involve solving a quartic equation instead. To avoid this problem, we propose an alternative representation of the lens shape. 
We consider the general case that the  front surface is fixed and is a height field over a domain $U$ in the $x$-$y$ space with a parametric form $z = \frontheightfunc(x,y)$, and the back surface is a triangle mesh. Additionally, we assume the following:
\begin{description}
\item[(A1)]  The rays refracted at the front surface define a one-to-one correspondence $\frontbackmap$ from the front surface to the back surface.
\item[(A2)] The front surface has no self-occlusion w.r.t. the light source.
\item[(A3)]  The direction vector of any refracted light ray has a positive $z$-coordinate.
\end{description}

\begin{wrapfigure}{r}[1em]{0.4\columnwidth}
\hspace*{-3em}
\centering
\includegraphics[width=0.39\columnwidth]{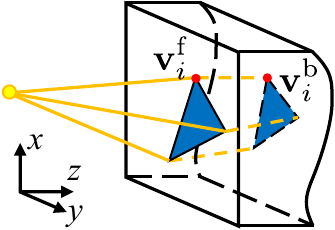}
\end{wrapfigure}
Based on Assumption~(A1), the back surface mesh induces a triangulation 
on the front surface, such that each vertex $\frontvtx_i$ of the front surface corresponds to a vertex $\backvtx_i = \frontbackmap(\frontvtx_i)$ on the back surface (see inset figure). 
Let $\zback_i$ be the $z$-coordinate of $\backvtx_i$, and $(\xfront_i, \yfront_i)$ be the $x$- and $y$-coordinates of $\frontvtx_i$. Then using the parametric form of the front surface, we can derive the position and the normal of  $\frontvtx_i$ as
\[
    \frontvtx_i = ( \xfront_i, \yfront_i, \frontheightfunc(\xfront_i, \yfront_i)),
    \quad
    \frontnormal_i = (-\frontheightfunc_x(\xfront_i, \yfront_i), -\frontheightfunc_y(\xfront_i, \yfront_i), 1).
\]
The incident light direction at $\frontvtx_i$ is $\incomingdir_1^{\text{f}} = \frontvtx_i - \pointsource $ where $\pointsource$ is the light source position. We can then compute the refracted light direction $\outgoingdir_i^{\text{f}}$ at $\frontvtx_i$ using Eq.~\eqref{eq:RefractedDir}. 
By definition, $\outgoingdir_i^{\text{f}}$ is also the incident light direction at the back surface vertex $\backvtx_i$. Moreover, as $\frontvtx_i$ lies on the ray emitted from $\frontvtx_i$ along direction $\outgoingdir_i^{\text{f}}$, there exists a value $t_i$ with 
\begin{equation}
\backvtx_i = \frontvtx_i + t_i \outgoingdir_i^{\text{f}}.
\label{eq:backvtxpos}
\end{equation}
Based on Assumption~(A3), it is easy to know that 
\begin{equation}
    t_i = \frac{\zback_i -  \frontheightfunc(\xfront_i, \yfront_i)}{z_i^{\text{r}}}, 
    \label{eq:tvalue}
\end{equation}
where $z_i^{\text{r}}$ is the z-coordinate of  $\outgoingdir_i^{\text{f}}$. Substituting \eqref{eq:tvalue} into \eqref{eq:backvtxpos}, we obtain the coordinates of $\backvtx_i$.
To summarize, the position and incident light direction for each back surface vertex $\backvtx_i$ can be parameterized by its $z$-coordinate together with the $(x,y)$-coordinates of its corresponding front surface vertex $\frontvtx_i$. Therefore, we use these parameters as variables to encode the lens shape and perform optimization. 

It is worth noting that the mesh representation in Sec.~\ref{sec:optimization} is a special case of the representation described here. In particular, with a parallel light source, the front surface vertices have the same  $(x,y)$-coordinates as their corresponding back surface vertices. Therefore, the variables in our generalized representation are exactly the coordinates of the back surface vertices, which is the same as in Sec.~\ref{sec:optimization}.

Besides the representation, the calculation of flux for a back surface triangle must also be changed when using a point light source.
We assume the light source to be Lambertian (i.e., with the same intensity along different directions). Then from Assumption~(A2) we can compute the flux for a back surface triangle $\backtriangle_i$ as:
\[
    \Phi_i = \frac{\solidangle(\fronttriangle_i)}{\sum\nolimits_{j=1}^{n_t}\solidangle(\fronttriangle_j)},
\]
where $\fronttriangle_i$ is the corresponding front surface triangle for $\backtriangle_i$, and $\solidangle(\cdot)$ denotes the solid angle w.r.t. the light source.

\section{Results}
\label{sec:experiment}

In this section, we verify the effectiveness of our algorithm, and showcase physical prototypes to demonstrate its practicality. All results in this paper are computationally simulated except for those explicitly identified as physical prototype photographs.

\paraheading{Implementation details}
The experimental settings for the examples presented in this paper are summarized in Table \ref{table:parameters}. 
For the example in the last row, the initial reflective surface has an incident angle of $30^{\circ}$ and is parallel to the receptive plane.
The weights for Eqs.~\eqref{eq:TriangleOptimization} and \eqref{eq:MeshUpdateProblem} are reported in Appendix~\ref{appx:details}.

\begin{table}[t]
\caption{Experimental settings of our results, and the MAE compared to the target image. The results in the last two rows are based on refraction of a point light source and reflection of parallel light respectively, while the other results are based on refraction of parallel light.}
\centering
\setlength{\tabcolsep}{1.3pt}
\begin{tabular}{ccccccc}
\toprule
\multirow{3}{*}{Image} & \multirow{3}{*}{\#Pixels} & \multirow{3}{*}{\#Vertices} & Lens & Image & Focal & MAE \\
  & & & Size & Size & Length & ($\times$$10^{-3}$)\\
  & & & (cm) & (cm) & (cm) & \\
\midrule 
SIGGRAPH & $512^2$ & $641$$\times$$737$ & $10$$\times$$10$$\times$$1.7$ & $10$$\times$$10$ & $30$ & 1.029 \\
Einstein & $512^2$ & $641$$\times$$737$ & $10$$\times$$10$$\times$$1.7$ & $10$$\times$$10$ & $30$ & 2.128\\
Train & $256^2$ & $641$$\times$$737$ & $10$$\times$$10$$\times$$1.6$ & $10$$\times$$10$ & $30$ & 3.470\\
Butterfly & $256^2$ & $641$$\times$$737$ & $10$$\times$$10$$\times$$1.0$ & $10$$\times$$10$ & $30$ & 0.054\\
Calligraphy & $800$$\times$$400$ & $833$$\times$$481$ & $20$$\times$$10$$\times$$1.4$ & $20$$\times$$10$ & $40$ &  6.235\\
Newton & $256^2$ & $641$$\times$$737$ & $10$$\times$$10$$\times$$0.8$ & $10$$\times$$10$ & $30$ & 0.001 \\
Butterfly (pt.) & $256^2$ & $641$$\times$$737$ & $10$$\times$$10$$\times$$0.9$ & $20$$\times$$20$ & $60$ & 1.586 \\
Newton (refl.) & $256^2$ & $641$$\times$$737$ & $10$$\times$$10$$\times$$0.7$ & $10$$\times$$10$ & $60$ & 0.000\\
\bottomrule
\label{table:parameters}
\end{tabular}
\end{table}

\subsection{Results of our method}

\begin{figure}[t]
\centering
\includegraphics[width=1\linewidth]{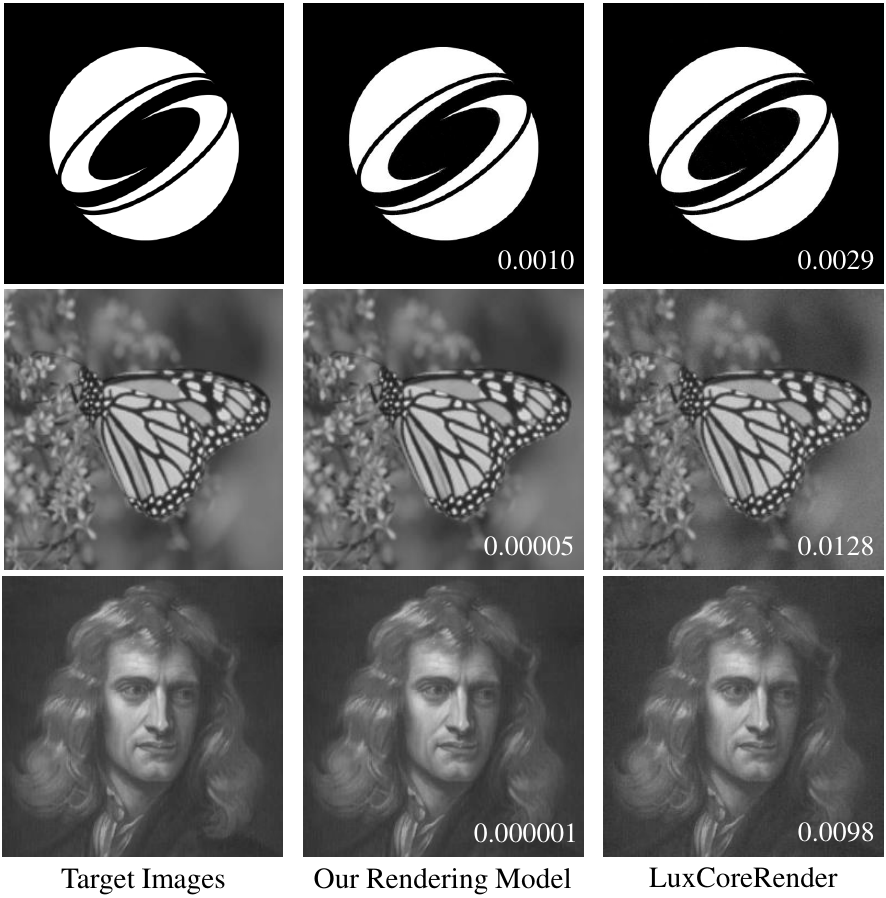}
\caption{Accuracy of our rendering model. The three examples from top to bottom are based on refraction of parallel light, refraction of point light, and reflection of parallel light, respectively. The number in the lower right corner of each image represents the MAE with the target image. Newton portrait courtesy of Isaac Newton Institute for Mathematical Sciences, University of Cambridge.}
\label{fig:Blender}
\end{figure}

\paraheading{Accuracy of our rendering model}
In Fig.~\ref{fig:Blender}, we validate the accuracy of our triangle-based rendering by comparing our rendered results of the final image with those generated by the physically-based renderer \luxcorerender\footnote{\url{https://luxcorerender.org/}}, using examples of refraction (with both parallel and point light sources) and reflection. It is evident that our triangle-based rendering closely resembles the ray-traced rendering in both scenarios, confirming the reliability of our rendering model.
In contrast to \luxcorerender, our approach achieves real-time rendering by computing only the position of each triangle, without the need for complex sampling techniques to trace individual light rays. Additionally, our rendering model is differentiable, allowing seamless integration into our end-to-end optimization.

In Fig.~\ref{fig:Blender}, we also evaluate the difference between the rendered images and the target image using their \emph{mean absolute error} (MAE):
\begin{equation}
\mae = \frac{1}{n_p} \sum_{i=1}^{n_p} \pixelerr_i,
\label{eq:MAE}
\end{equation}
where $\pixelerr_i$ is the pixel-wise absolute error defined as:
\begin{equation}
\pixelerr_i = \frac{| \pixel{i} - \pixeltwo{i}|}{\pixelmax},
\label{eq:pixelerr}
\end{equation}
with $\pixel{i}, \pixeltwo{i}$ being corresponding pixels from the images to be compared, and $\pixelmax$ being the maximum possible pixel value. The MAE values show that regardless of the rendering method used, our resulting image is very close to the target image.

\begin{figure*}[t]  
\begin{minipage}{0.02\linewidth}
\rotatebox{90}{Target Image}
\end{minipage}
\hfill
\begin{minipage}{0.152\linewidth}
  \centerline{\includegraphics[width=1\linewidth]{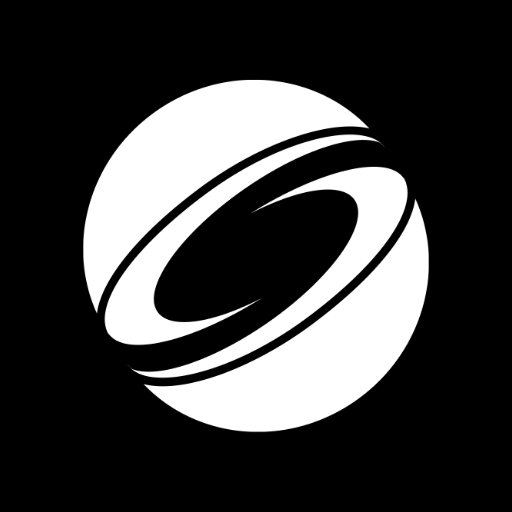}}
\end{minipage}
\hfill
\begin{minipage}{0.152\linewidth}
  \centerline{\includegraphics[width=1\linewidth]{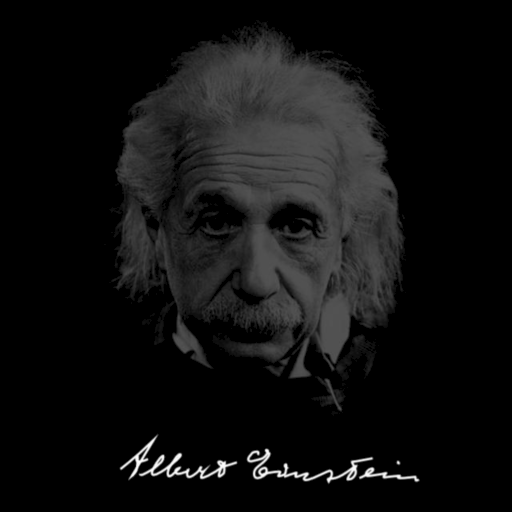}}
\end{minipage}
\hfill
\begin{minipage}{0.152\linewidth}
  \centerline{\includegraphics[width=1\linewidth]{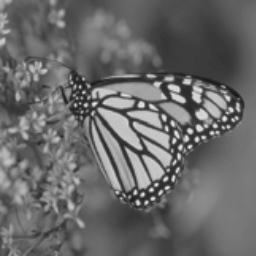}}
\end{minipage}
\hfill
\begin{minipage}{0.152\linewidth}
  \centerline{\includegraphics[width=1\linewidth]{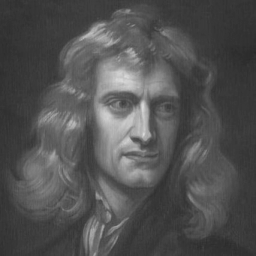}}
\end{minipage}
\hfill
\begin{minipage}{0.304\linewidth}
  \centerline{\includegraphics[width=1\linewidth]{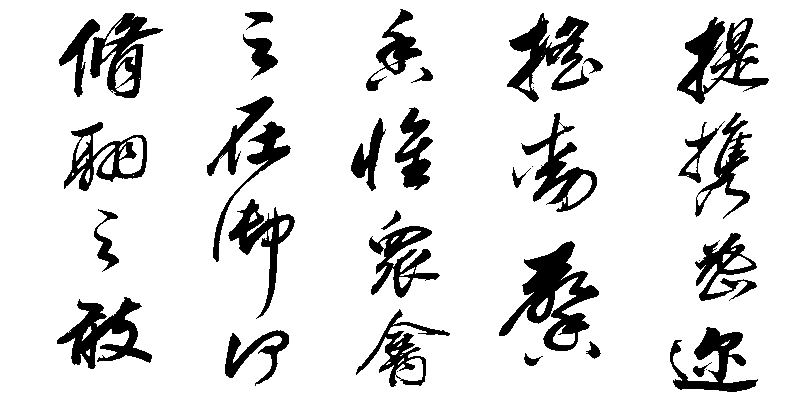}}
\end{minipage}
\hfill
\hspace{0.025\linewidth}
\vfill
\vspace{0.2em}
\begin{minipage}{0.02\linewidth}
\rotatebox{90}{Rendered Image}
\end{minipage}
\hfill
\begin{minipage}{0.152\linewidth}
  \centerline{\includegraphics[width=1\linewidth]{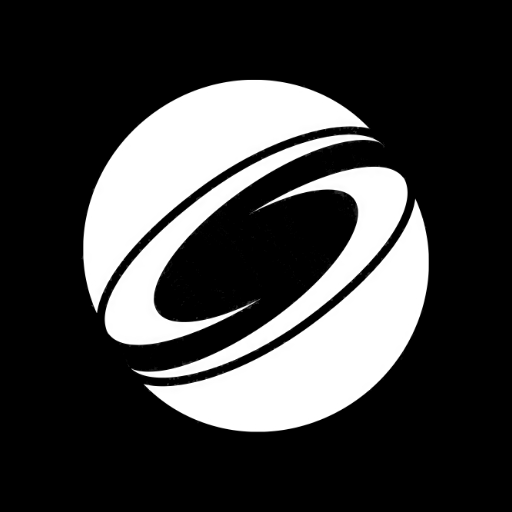}}
\end{minipage}
\hfill
\begin{minipage}{0.152\linewidth}
  \centerline{\includegraphics[width=1\linewidth]{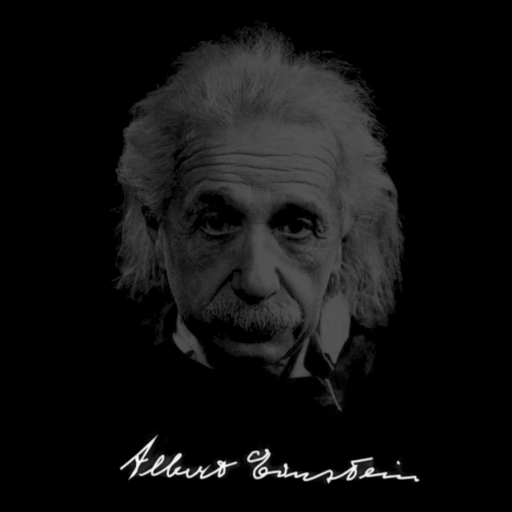}}
\end{minipage}
\hfill
\begin{minipage}{0.152\linewidth}
  \centerline{\includegraphics[width=1\linewidth]{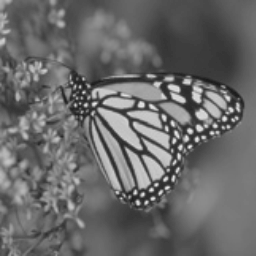}}
\end{minipage}
\hfill
\begin{minipage}{0.152\linewidth}
  \centerline{\includegraphics[width=1\linewidth]{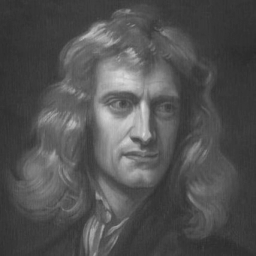}}
\end{minipage}
\hfill
\begin{minipage}{0.304\linewidth}
  \centerline{\includegraphics[width=1\linewidth]{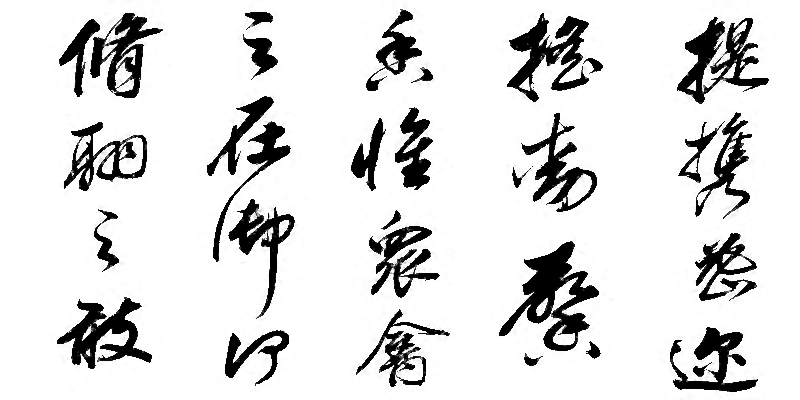}}
\end{minipage}
\hfill
\hspace{0.025\linewidth}
\vfill
\vspace{0.2em}
\begin{minipage}{0.02\linewidth}
\rotatebox{90}{Error}
\end{minipage}
\hfill
\begin{minipage}{0.152\linewidth}
  \centerline{\includegraphics[width=1\linewidth]{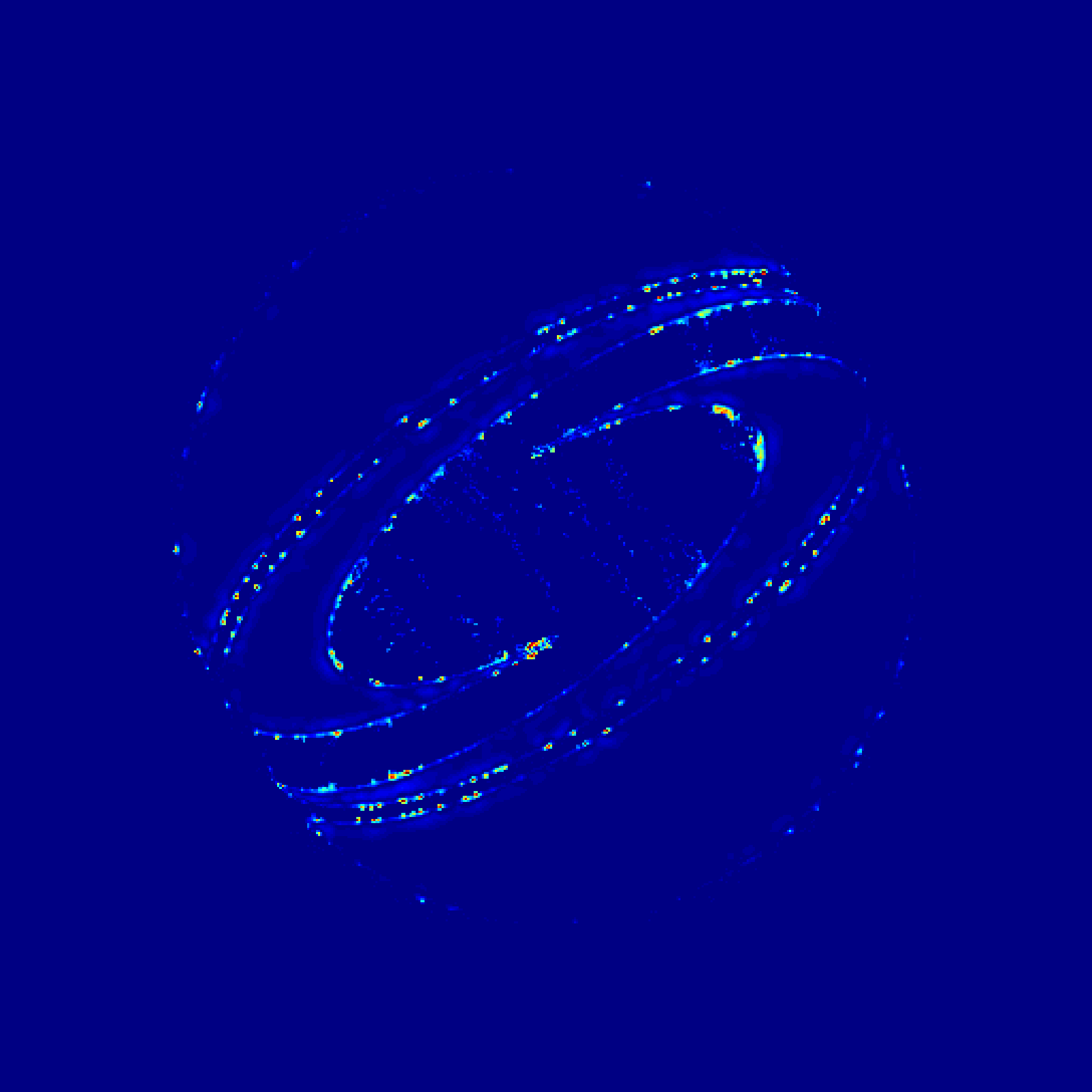}}
\end{minipage}
\hfill
\begin{minipage}{0.152\linewidth}
  \centerline{\includegraphics[width=1\linewidth]{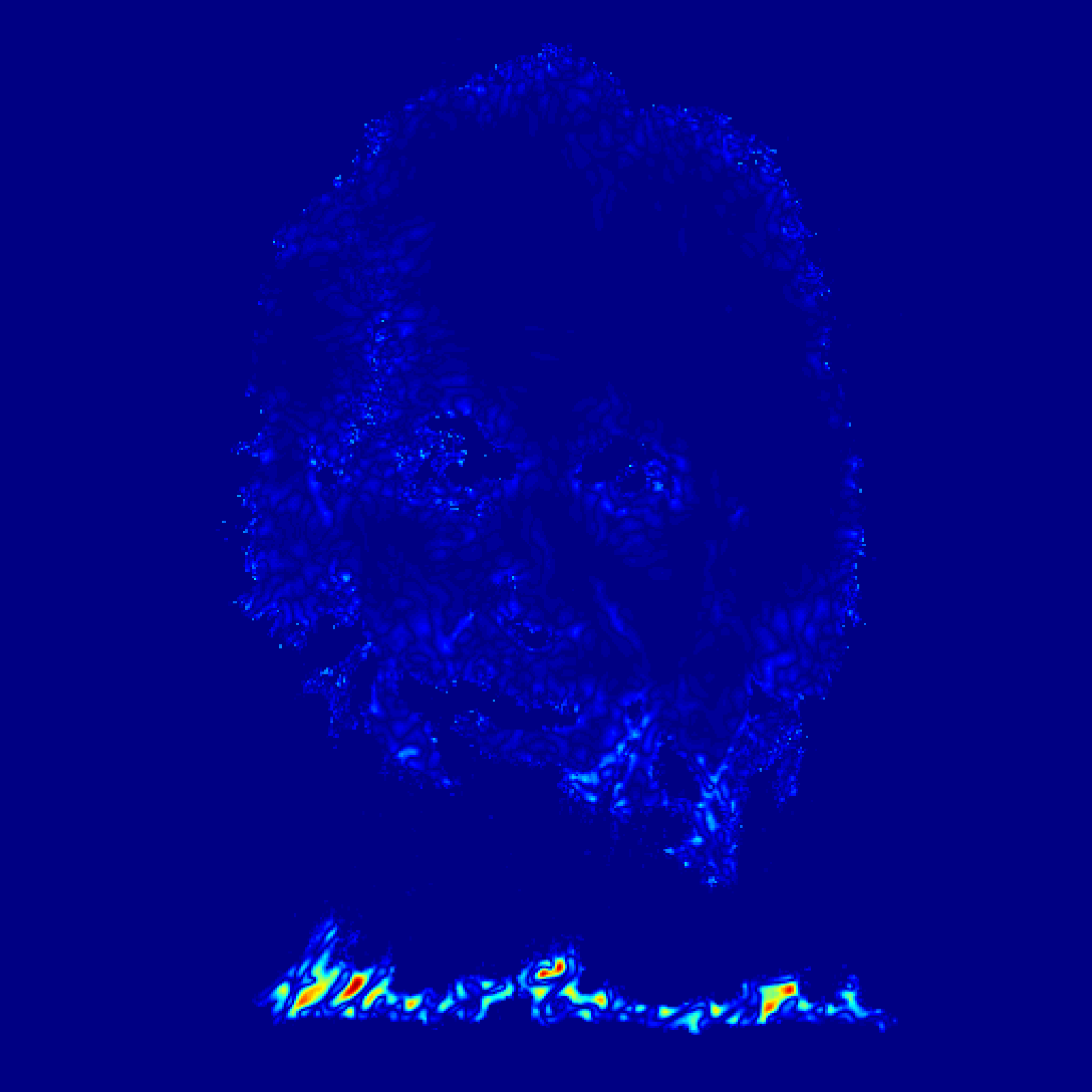}}
\end{minipage}
\hfill
\begin{minipage}{0.152\linewidth}
  \centerline{\includegraphics[width=1\linewidth]{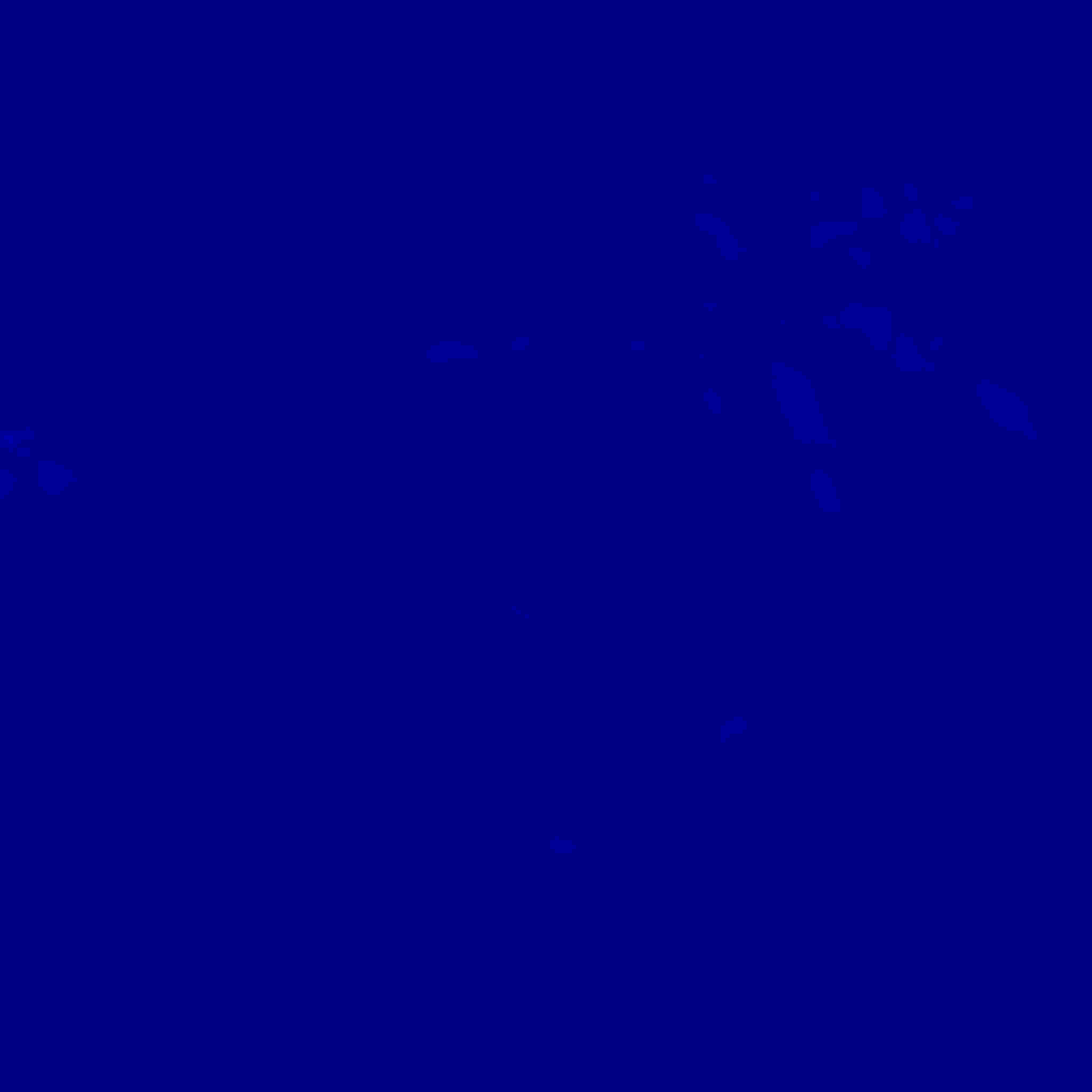}}
\end{minipage}
\hfill
\begin{minipage}{0.152\linewidth}
  \centerline{\includegraphics[width=1\linewidth]{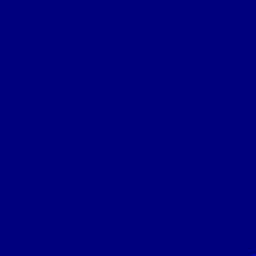}}
\end{minipage}
\hfill
\begin{minipage}{0.304\linewidth}
  \centerline{\includegraphics[width=1\linewidth]{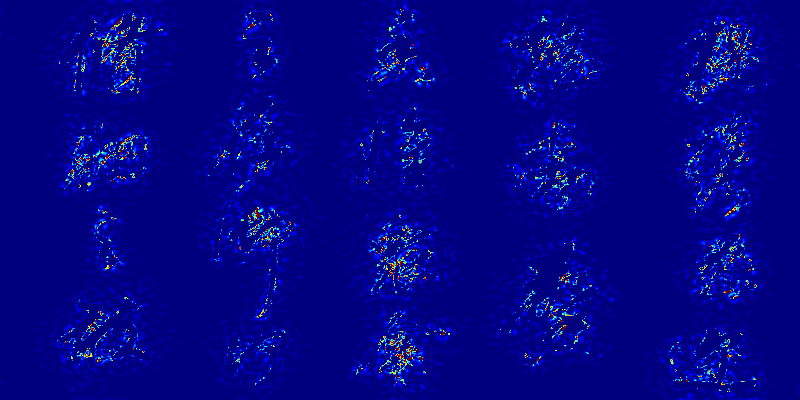}}
\end{minipage}
\hfill
\begin{minipage}{0.025\linewidth}
  \centerline{\includegraphics[width=1\linewidth]{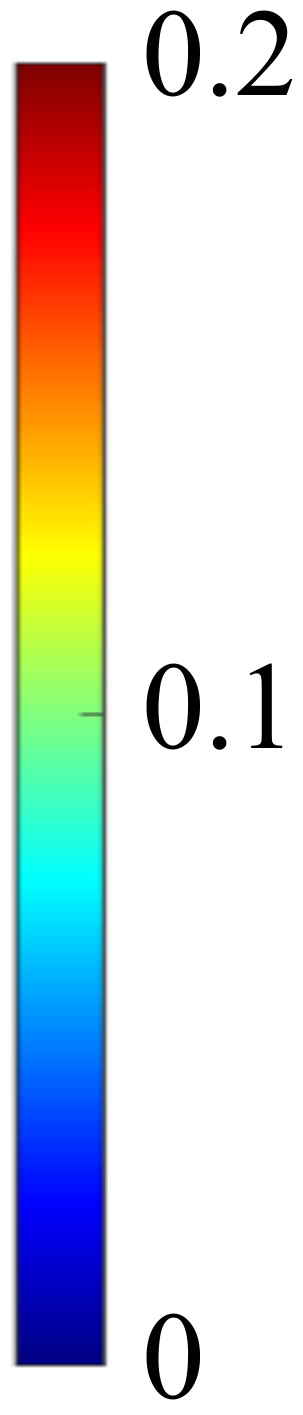}}
\end{minipage}
\caption{Our optimization produces results with their rendered image very close to the target images. From top to bottom: target images, rendered resulting images of our lens, and color-coding of pixel-wise absolute errors defined in Eq.~\eqref{eq:pixelerr}. Einstein portrait \textcopyright~Magnum/IC photo; Newton portrait courtesy of Isaac Newton Institute for Mathematical Sciences, University of Cambridge; Calligraphy image source: National Palace Museum, Taipei, CC BY 4.0.}
\label{fig:colorbar}
\end{figure*}

\paraheading{Effectiveness of our optimization}
We evaluate the effectiveness of our optimization approach by comparing the target image with the resulting image generated using our rendering model, measuring their MAE. The MAE values for different examples are shown in the last column of Table~\ref{table:parameters}. Furthermore, Fig.~\ref{fig:colorbar} visualizes the pixel errors for some examples using color-coding. Both the MAE values and the pixel error maps confirm that our method successfully computes lens shapes that produce images nearly identical to the target.

\subsection{Effectiveness of algorithmic components}
\begin{figure*}[t!]
\centering
\includegraphics[width=1.0\linewidth]{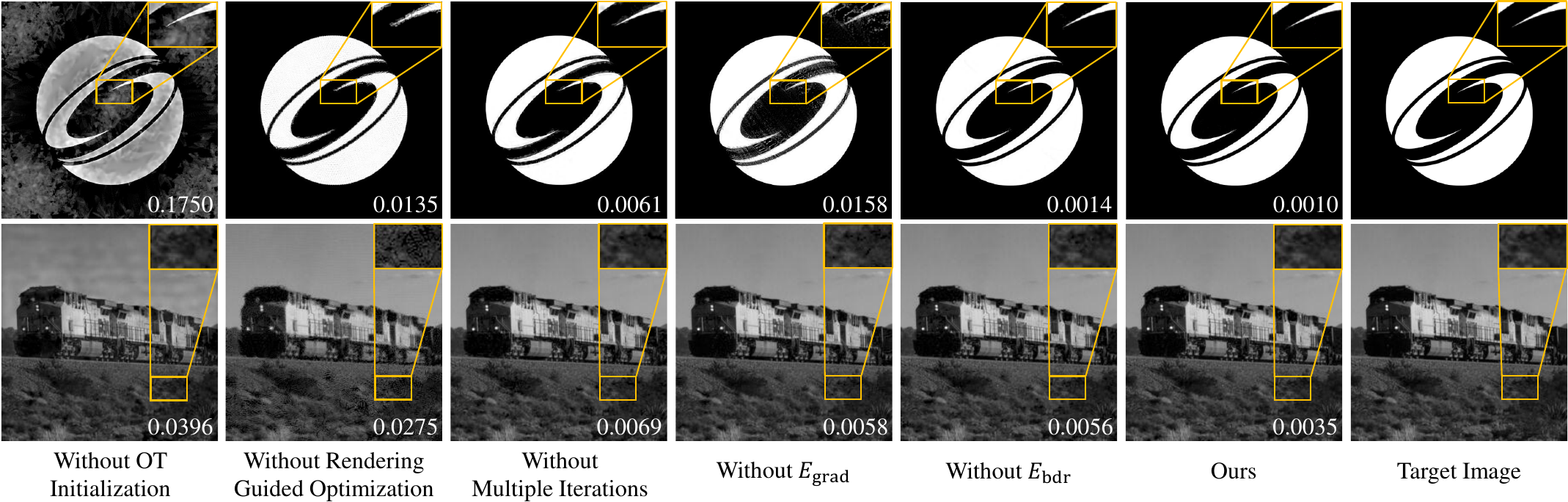}\\
\vspace{-0.5em}
\caption{Comparison between different variants of our method, with some algorithmic components removed. The number in the lower right corner of each image represents the MAE with the target image.}
\label{fig:ablation_study}
\end{figure*}

In Fig.~\ref{fig:ablation_study}, we conduct controlled experiments to verify the effectiveness of each component in our algorithm.

The column ``Without OT Initialization'' in Fig.~\ref{fig:ablation_study} highlights the importance of initialization. For high-contrast images such as the SIGGRAPH logo, it is challenging to rely solely on image errors to move the light from a black region to a distant area where it is needed. For the grayscale image, although a simple plane initialization without OT achieves a satisfactory result, there are still slight tone discrepancies compared to the target image, and wrinkled artifacts that affect the overall visual quality.

To verify the necessity of rendering guided optimization, we attempt iterative optimization using only OT initialization. The results are shown in the column ``Without Rendering Guided Optimization'' of Fig.~\ref{fig:ablation_study}. Although this approach achieves the black background for the SIGGRAPH logo, it fails to recover the sharp boundaries of the white regions fully. For the grayscale image, undesirable black gaps emerge in some parts of the image. Both issues are due to the lack of rendering guided optimization, designed to preserve fine details from the target image.

In the column ``Without Multiple Iterations'', we investigate the impact of iterative optimization by only performing a single round of OT followed by six rendering-based optimization rounds (i.e., using only one OT while significantly increasing the iteration count of rendering-based optimization) for each mesh resolution. 
This leads to higher MAE values and lower quality of the resulting image, because an optimization trajectory may encounter multiple local minima, which requires more than a single OT correction to arrive at a desirable solution.
In comparison, our approach of alternating between these two procedures (as shown in the column ``Ours'') can fully leverage their strengths, continually updating and refining the current solution, allowing us to achieve a rendered image almost identical to the target.

\begin{figure}[t]
\centering
\includegraphics[width=0.97\linewidth]{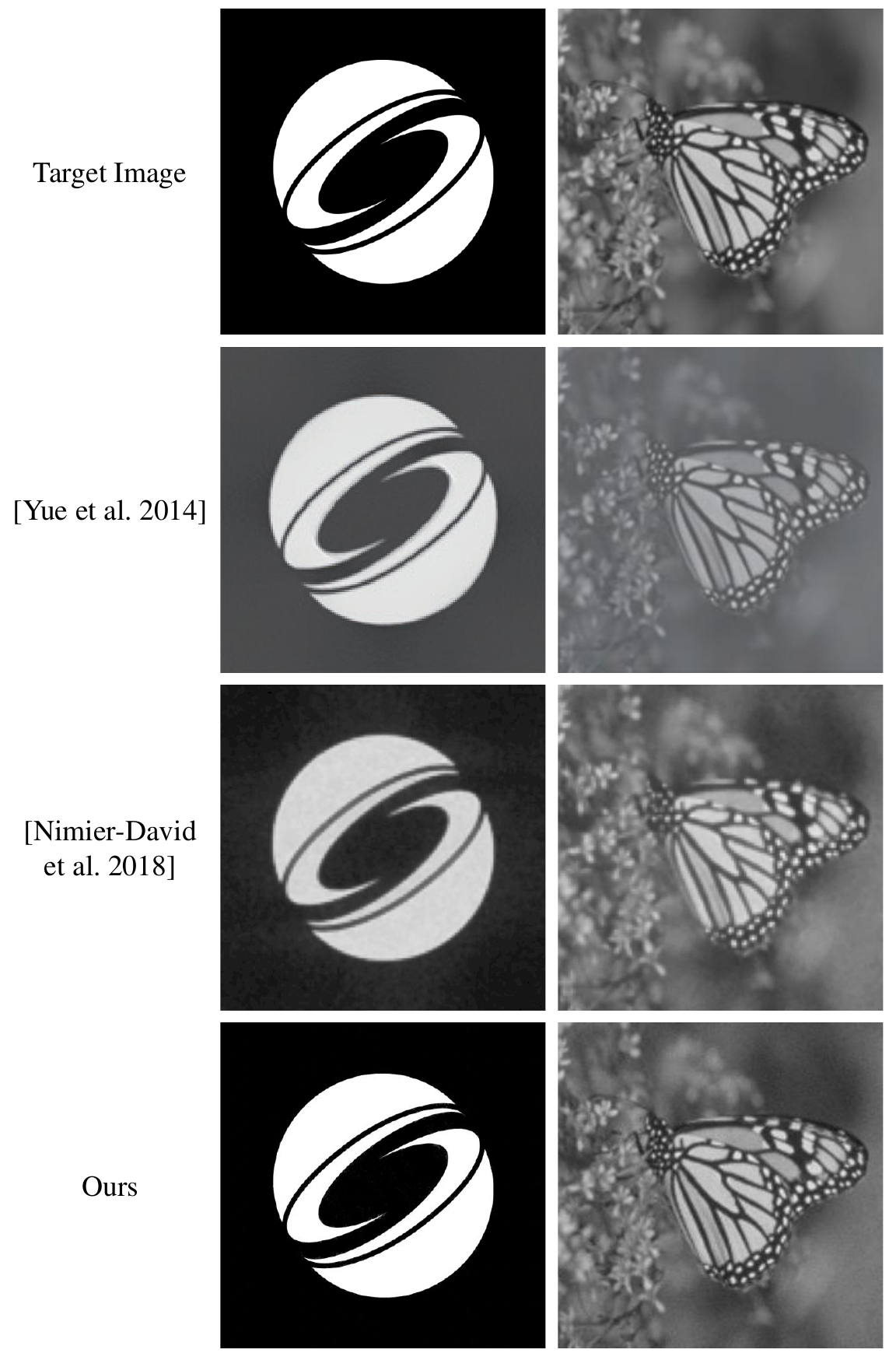}
\vspace{-0.7em}
\caption{Comparison with \cite{yue2014poisson} and \cite{NimierDavidVicini2019Mitsuba2}. The results of~\cite{yue2014poisson} are taken from their paper, while the results of \cite{NimierDavidVicini2019Mitsuba2} are generated using Mitsuba~3.}
\label{fig:compare_yue_mitsuba}
\end{figure}

We also verify the necessity of the target function terms in our optimization formulation~\eqref{eq:TriangleOptimization}. In the column ``Without $\gradloss$'', we perform the optimization without the term $\gradloss$. This results in a weaker capability to preserve image features and causes an increase in MAE, which is particularly noticeable in high-contrast areas. 
In the column 
``Without $\boundaryloss$'', we remove the term $\boundaryloss$ from the target function, which also results in higher MAE values. 
This is because the lack of $\boundaryloss$ allows the optimizer to move some lights out of the imaging area in order to reduce other non-image terms in the target function, which can cause deviation from the target image.
Finally, later in Sec.~\ref{sec:PhysicalPrototypes} and Fig.~\ref{fig:prototype_smooth}, we also demonstrate the necessity of the term $\fullsmoothloss$ for ensuring the quality of the physical results.

\subsection{Comparison with existing methods}
\begin{figure}[t]
\centering
\includegraphics[width=0.97\linewidth]{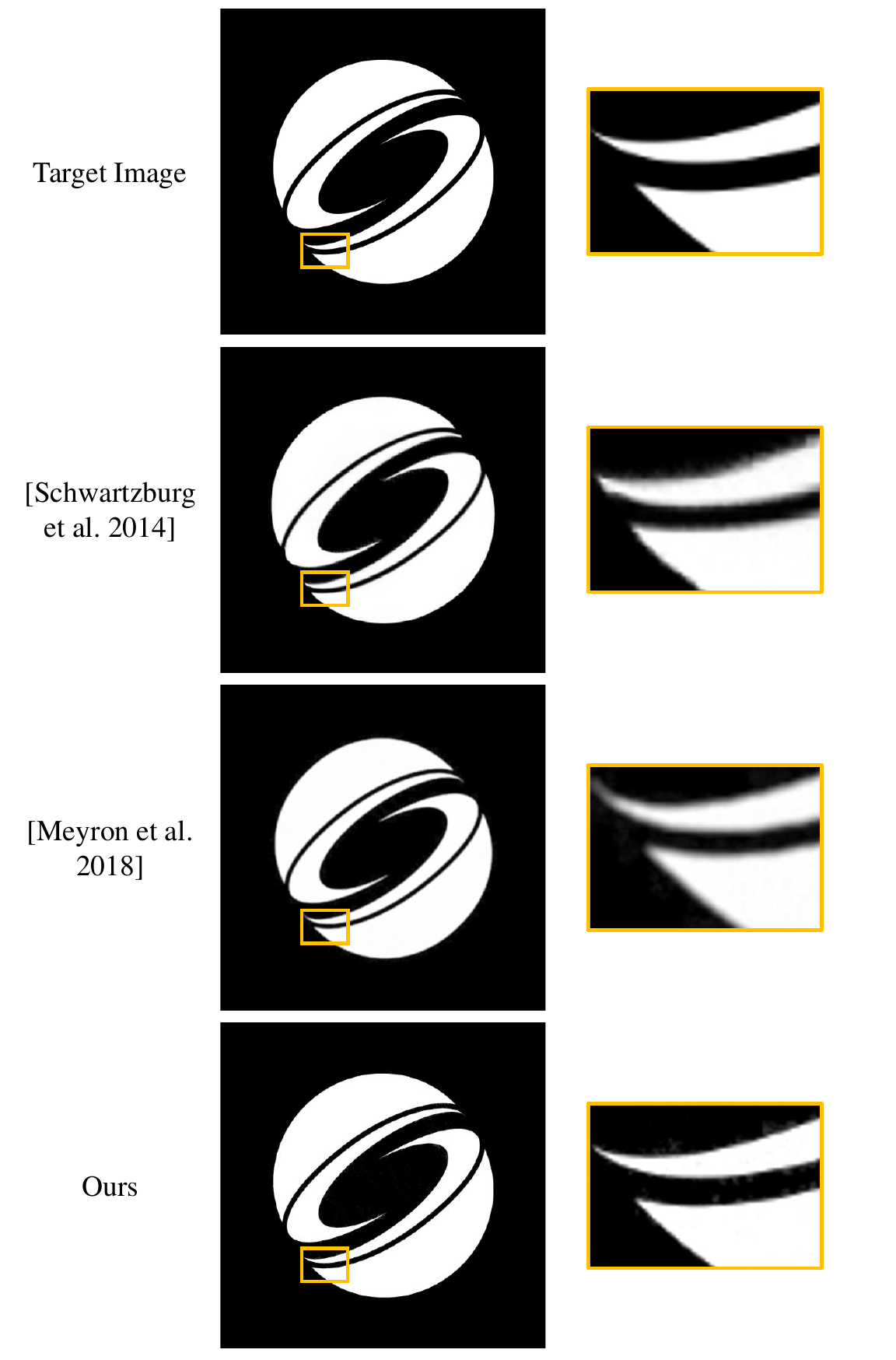}
\vspace{-0.7em}
\caption{Comparison with \cite{schwartzburg2014high} and \cite{meyron2018light} using results from their papers. Some areas are enlarged to better show the details. }
\label{fig:compare_Yuliy_meyron}
\end{figure}

\begin{figure*}[t!]
\centering
\begin{minipage}{0.02\linewidth}
\rotatebox{90}{{\small Parallel Source}}
\end{minipage}
\begin{minipage}{0.16\linewidth}
  \centerline{\includegraphics[width=1\linewidth]{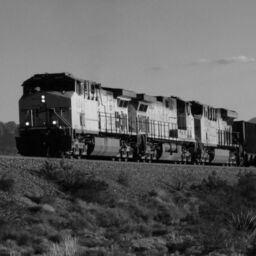}}
\end{minipage}
\addcolspace
\begin{minipage}{0.16\linewidth}
  \centerline{\includegraphics[width=1\linewidth]{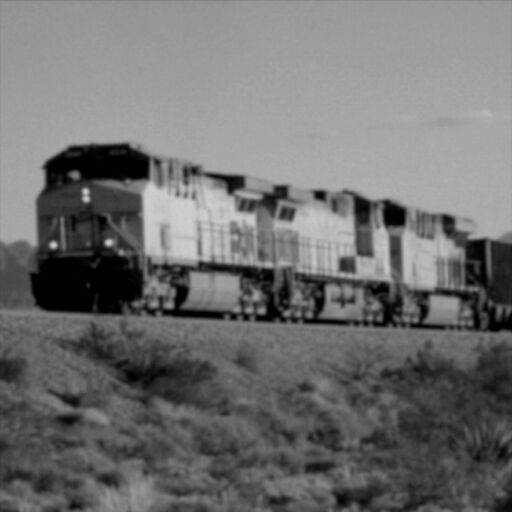}}
\end{minipage}
\begin{minipage}{0.16\linewidth}
  \centerline{\includegraphics[width=1\linewidth]{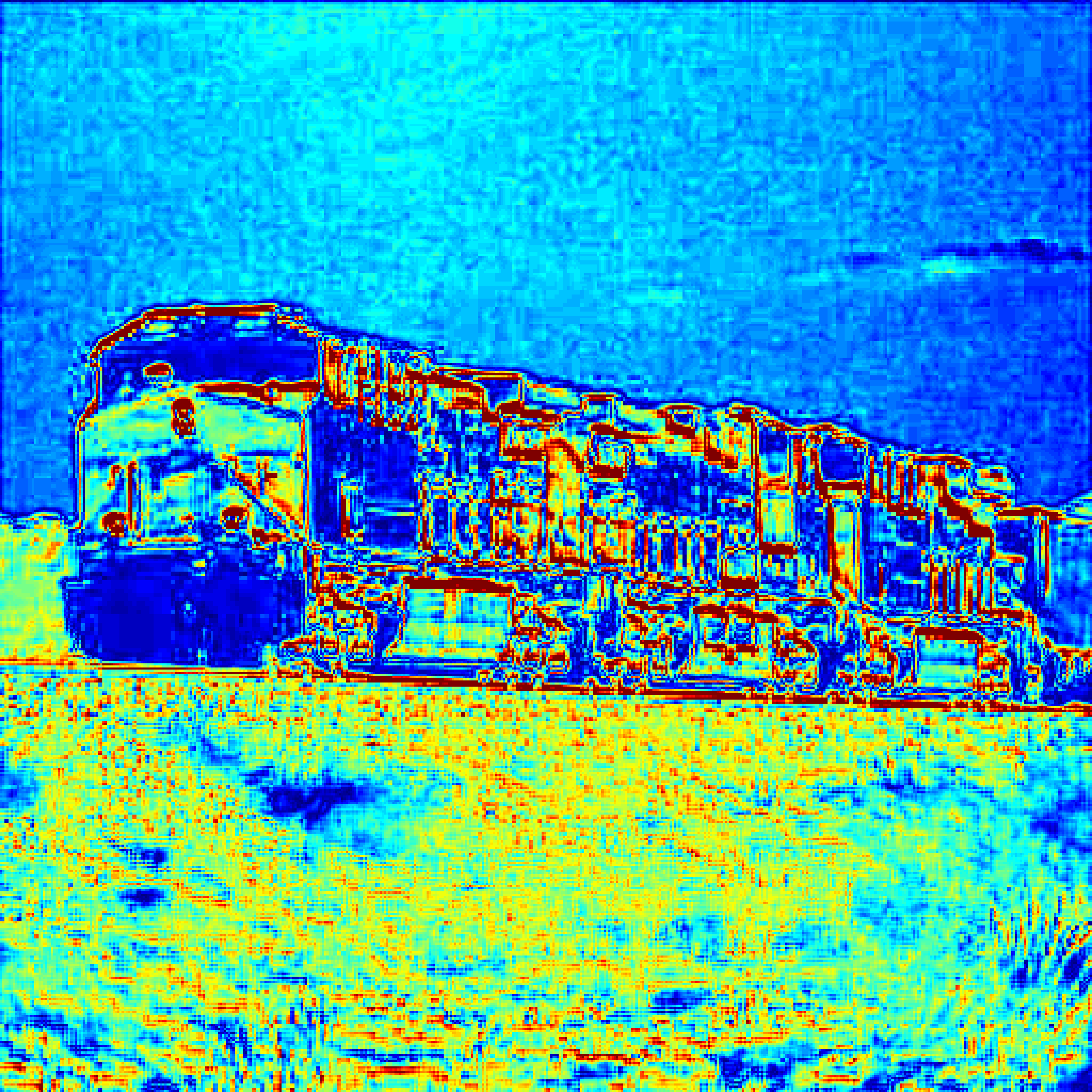}}
\end{minipage}
\addcolspace
\begin{minipage}{0.16\linewidth}
  \centerline{\includegraphics[width=1\linewidth]{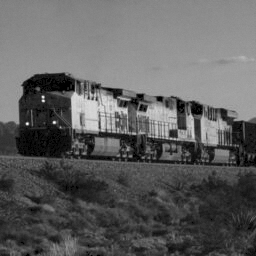}}
\end{minipage}
\begin{minipage}{0.16\linewidth}
  \centerline{\includegraphics[width=1\linewidth]{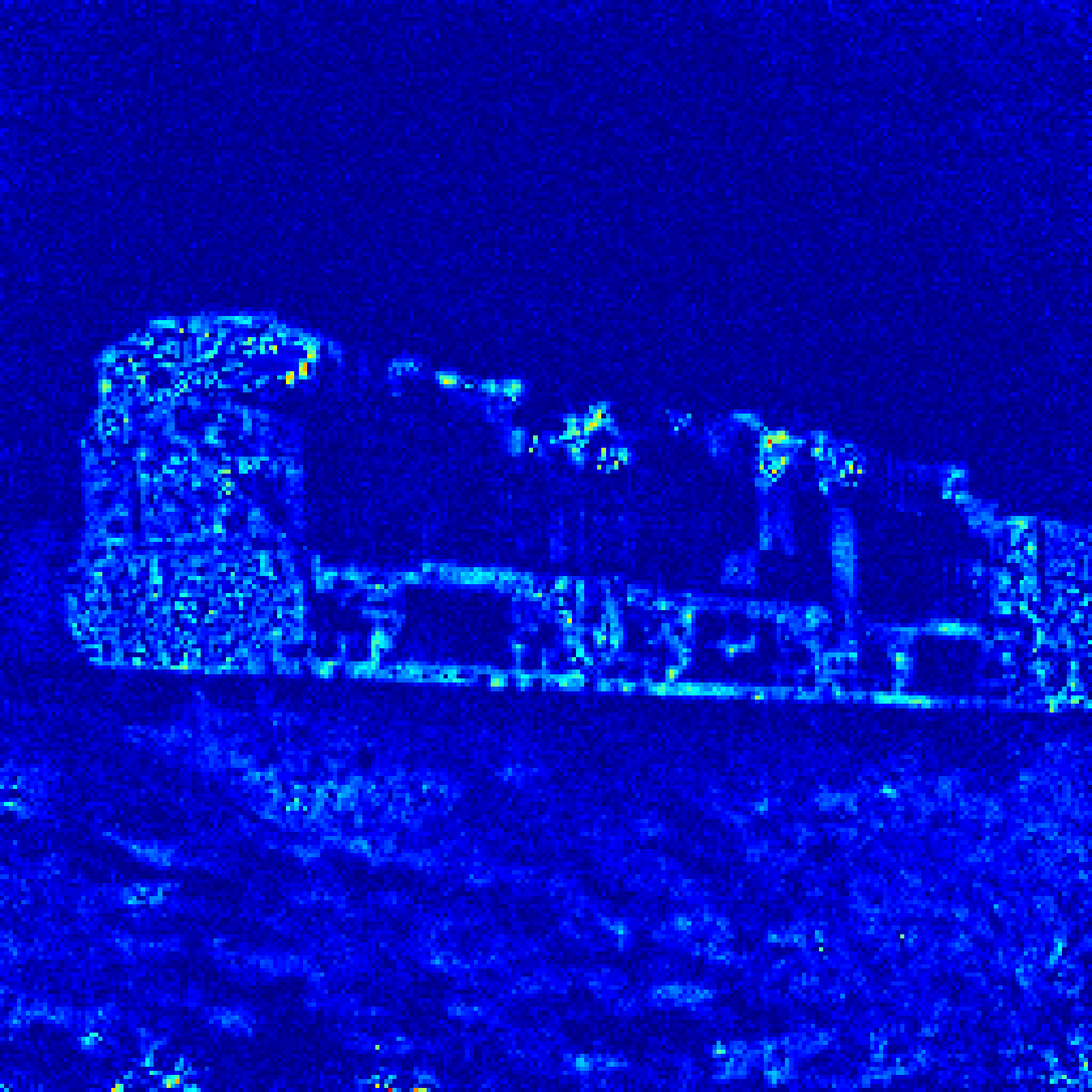}}
\end{minipage}
\hspace{0.2em}
\hspace{0.03\linewidth}
\vfill
\vspace{0.2em}
\begin{minipage}{0.02\linewidth}
\rotatebox{90}{{\small Point Source}}
\end{minipage}
\begin{minipage}{0.16\linewidth}
  \centerline{\includegraphics[width=1\linewidth]{Figs/Train_gt.png}}
\end{minipage}
\addcolspace
\begin{minipage}{0.16\linewidth}
  \centerline{\includegraphics[width=1\linewidth]{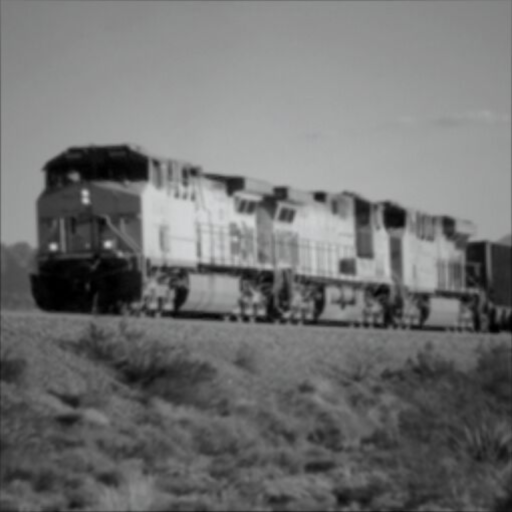}}
\end{minipage}
\begin{minipage}{0.16\linewidth}
  \centerline{\includegraphics[width=1\linewidth]{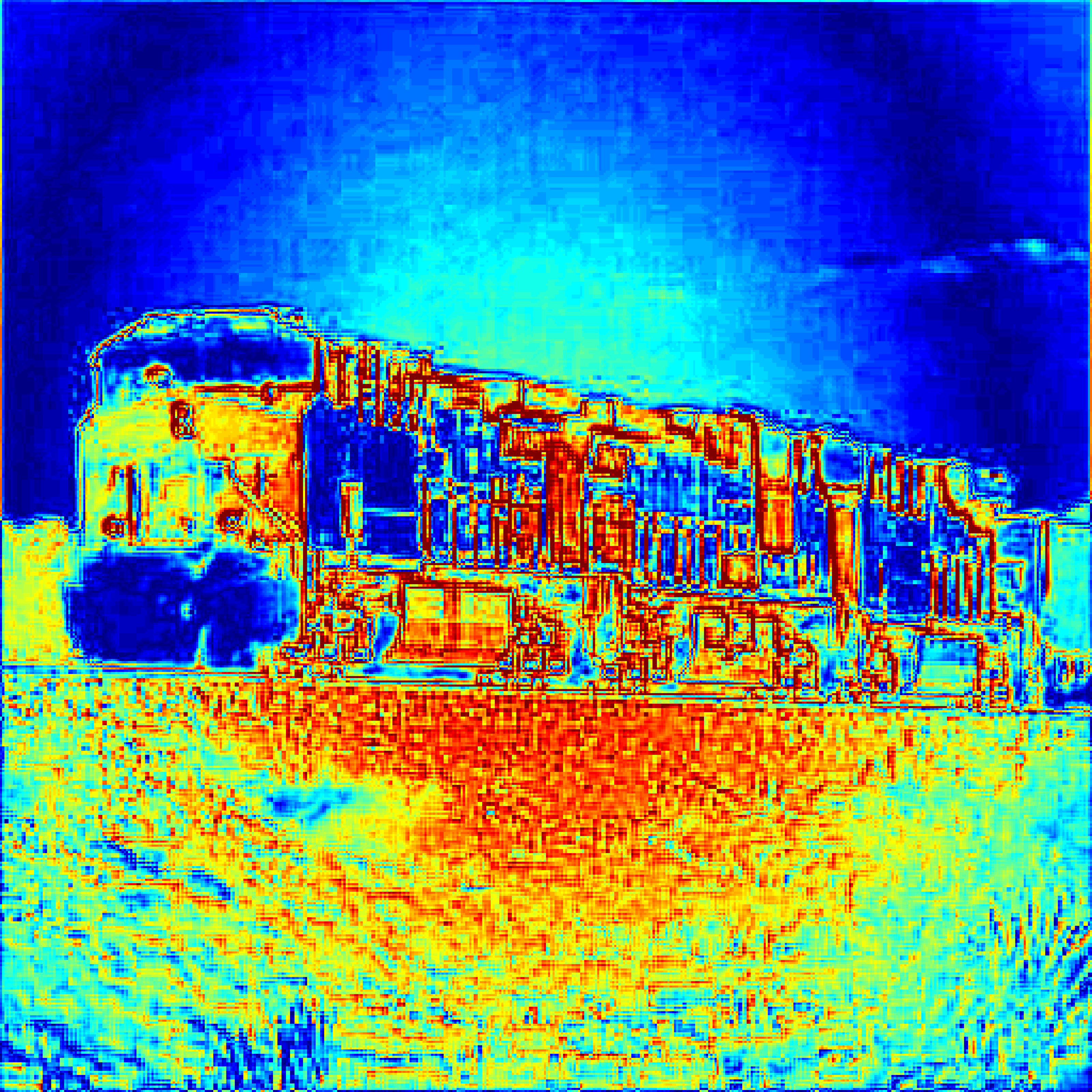}}
\end{minipage}
\addcolspace
\begin{minipage}{0.16\linewidth}
  \centerline{\includegraphics[width=1\linewidth]{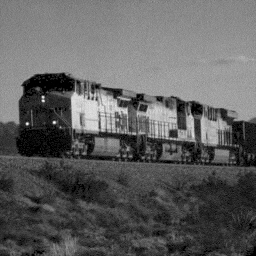}}
\end{minipage}
\begin{minipage}{0.16\linewidth}
  \centerline{\includegraphics[width=1\linewidth]{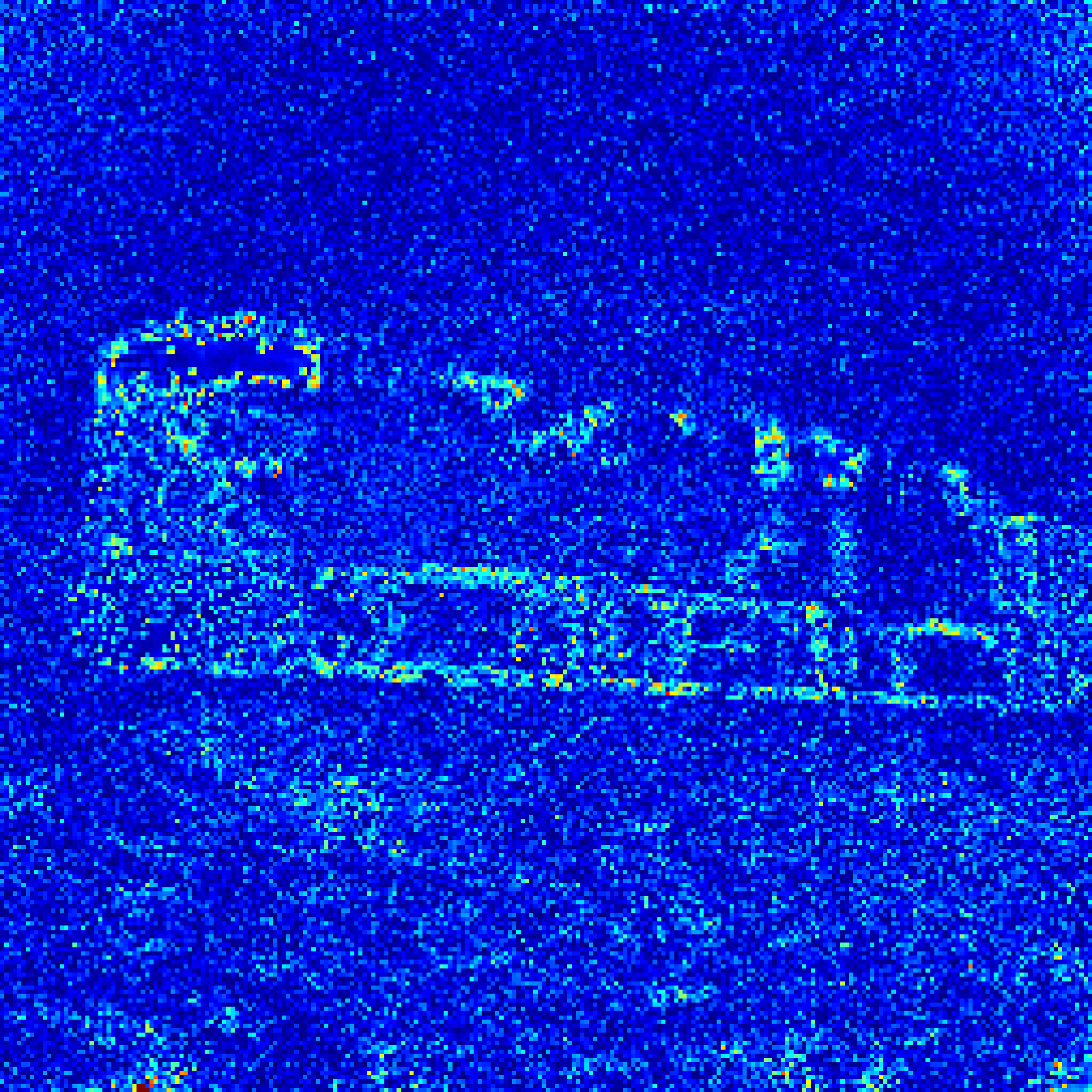}}
\end{minipage}
\hspace{0.2em}
\begin{minipage}{0.03\linewidth}
  \centerline{\includegraphics[width=1\linewidth]{Figs/Colorbar.pdf}}
\end{minipage}
\vfill
\vspace{0.4em}
\hspace{0.02\linewidth}
\begin{minipage}{0.16\linewidth}
  \centerline{{\small Target Image}}
\end{minipage}
\addcolspace
\begin{minipage}{0.32\linewidth}
  \centerline{{\small \cite{meyron2018light}}}
\end{minipage}
\addcolspace
\begin{minipage}{0.32\linewidth}
  \centerline{{\small Ours}}
\end{minipage}
\hspace{0.2em}
\hspace{0.03\linewidth}
\vspace{-0.7em}
\caption{Comparison with \cite{meyron2018light}. The color coding shows pixel-wise absolute error (Eq.~\eqref{eq:pixelerr}) compared to the target image.}
\label{fig:compare_meyron_train}
\end{figure*}

In Figs.~\ref{fig:compare_yue_mitsuba}, \ref{fig:compare_Yuliy_meyron} and \ref{fig:compare_meyron_train}, we compare our method with existing methods for caustics design, including~\cite{yue2014poisson}, \cite{schwartzburg2014high}, \cite{meyron2018light} and \cite{NimierDavidVicini2019Mitsuba2}, using the rendered images of the results. 
For a fair comparison, the images of our method are all generated using \luxcorerender{} rather than our rendering model.
The images of~\cite{yue2014poisson}, \cite{schwartzburg2014high} and \cite{meyron2018light} are taken from their papers, while the images of \cite{NimierDavidVicini2019Mitsuba2} are generated using Mitsuba~3\footnote{We follow the example given in \url{https://mitsuba.readthedocs.io/en/stable/src/inverse_rendering/caustics_optimization.html}.}.

Fig.~\ref{fig:compare_yue_mitsuba} compares our method with \cite{yue2014poisson} and \cite{NimierDavidVicini2019Mitsuba2}. Unlike our approach, both methods fail to produce rendered images with entirely black backgrounds effectively. For a grayscale target image, our method also achieves higher fidelity in the rendered image.

We also compare our work with \cite{schwartzburg2014high} and \cite{meyron2018light} in Fig.~\ref{fig:compare_Yuliy_meyron} using high-contrast target images. Although all three methods produce results of impressive quality, a closer examination reveals both \cite{schwartzburg2014high} and \cite{meyron2018light} produce narrower gaps between the two white regions than the target image. 
This is potentially due to accumulated errors through the process of first deriving intermediate features from the target image and then reconstructing the lens surface mesh from these features. Thanks to our differentiable rendering model and end-to-end optimization directly driven by the image difference, our approach can more effectively fine-tune the surface to match the target caustics, avoiding such discrepancies.
Fig.~\ref{fig:compare_meyron_train} further compares our method with~\cite{meyron2018light} on a grayscale target image using both a parallel light source and a point light source, and visualizes their pixel errors compared to the target. 
Our results are notably closer to the target under both light sources. The difference between the two methods is most notable in the lower part of the images where there are rich details. The results from~\cite{meyron2018light} smooth out some details and cause large pixel errors in the region. In contrast, our method can preserve the details thanks to the gradient term in our optimization formulation. 

A comprehensive quantitative comparison is presented in Table~\ref{table:comparison}. In addition to the MAE (Eq.~\eqref{eq:MAE}), we have also evaluated the Structural Similarity Index Measure (SSIM)~\cite{wang2004image}, which serves as a perceptual quality metric. SSIM measures the similarity between two images based on their structural information, luminance, and contrast. We use the target image as the reference for computing SSIM, with higher values indicating better perceived image quality and a maximum value of 1 corresponding to an exact match. The SSIM is calculated using the scikit-image library~\cite{scikit-image} with its default parameter values.
The results from~\cite{yue2014poisson} are not included in Table~\ref{table:comparison}, as we are unable to obtain the original result images to perform precise alignment with the target image for a fair comparison.
The experimental results demonstrate that our method significantly outperforms existing approaches across both evaluation metrics, indicating a closer match to the target image in terms of both pixel-wise accuracy and overall perceptual quality.

\begin{table}[t]
\caption{Quantitative comparison with existing methods using the MAE metric in Eq.~\eqref{eq:MAE} and the SSIM~\cite{wang2004image} with respect to the target image. The results in the last row are based on refraction of a point light source, while the others are based on refraction of parallel light.}
\centerline{MAE ($\times 10^{-3}$)$\downarrow$~~~~|~~~~SSIM$\uparrow$}
\centering
\setlength{\tabcolsep}{1.3pt}
\begin{tabular}{ccccc}
\toprule
\small{Image} & \parbox{40pt}{\small{\cite{schwartzburg2014high}}} & \parbox{35pt}{\small{\cite{meyron2018light}}} & \parbox{48pt}{\small{\cite{NimierDavidVicini2019Mitsuba2}}} & \small{Ours} \\
\midrule 
{\small SIGGRAPH} & 27.71 | 0.829 & / & 178.9 | 0.200 & \textbf{2.873 | 0.964}\\
{\small Einstein} & 25.10 | 0.773 & / & 64.56 | 0.124 & \textbf{12.32 | 0.840}\\
{\small Train} & / & 49.52 | 0.825 & 45.13 | 0.717 & \textbf{13.18 | 0.969}\\
{\small Train (pt.)} & / & 65.54 | 0.849 & / & \textbf{23.21 | 0.938}\\
\bottomrule
\label{table:comparison}
\end{tabular}
\end{table}

\subsection{Physical prototypes}
\label{sec:PhysicalPrototypes}

\begin{figure*}[t]
\centering
\begin{minipage}{0.1\linewidth}
\centering
\vfill
\small{Parallel Light}
\vfill
\end{minipage}
\addhspace
\begin{minipage}{0.4\linewidth}
\includegraphics[width=\linewidth]{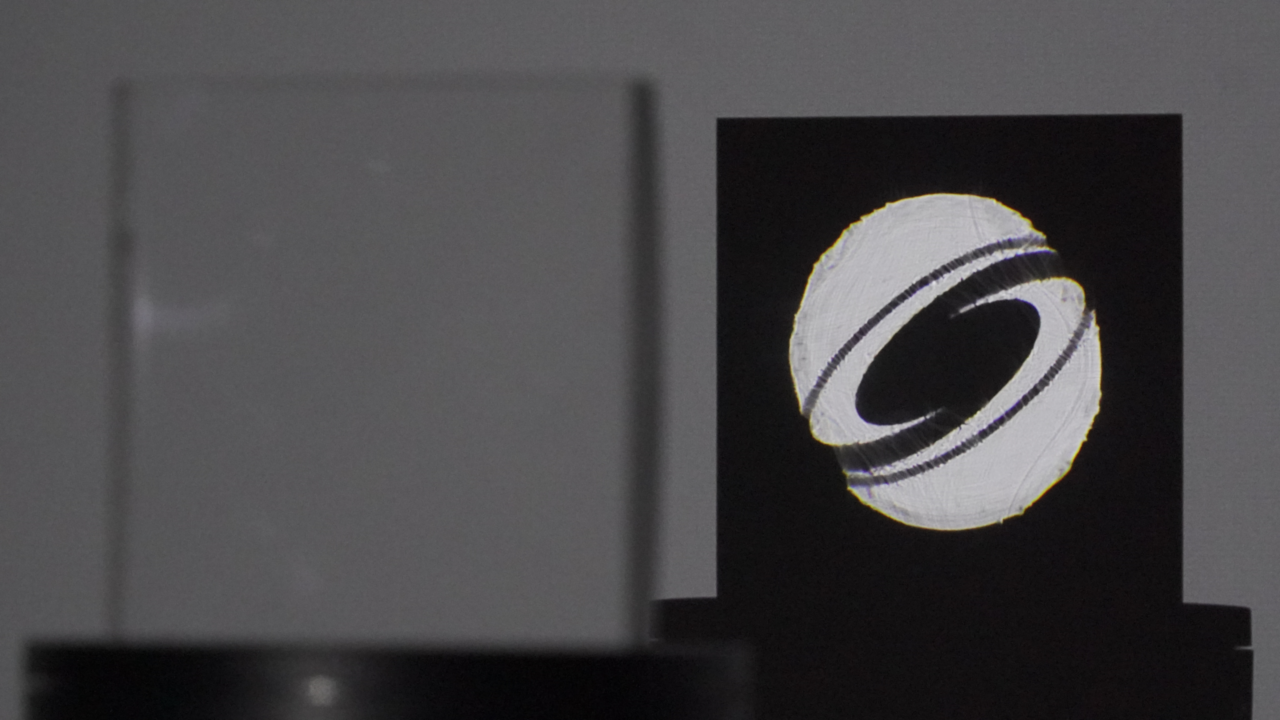}
\end{minipage}
\addhspace
\begin{minipage}{0.225\linewidth}
\includegraphics[width=\linewidth]{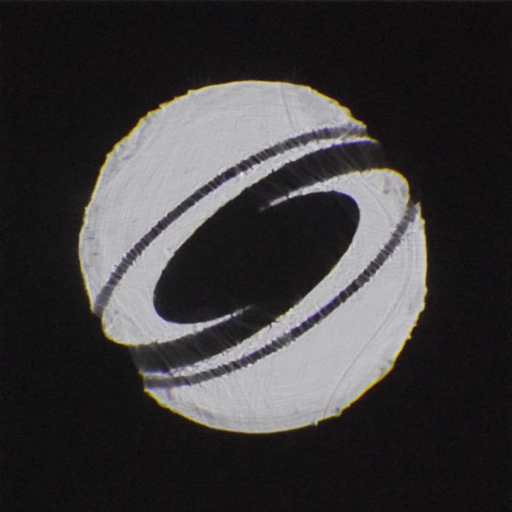}
\end{minipage}
\addhspace
\begin{minipage}{0.225\linewidth}
\includegraphics[width=\linewidth]{Figs/Siggraph_Ours.png}
\end{minipage}\\[0.4em]
\begin{minipage}{0.1\linewidth}
\centering
\vfill
\small{Parallel Light}
\vfill
\end{minipage}
\addhspace
\begin{minipage}{0.4\linewidth}
\includegraphics[width=\linewidth]{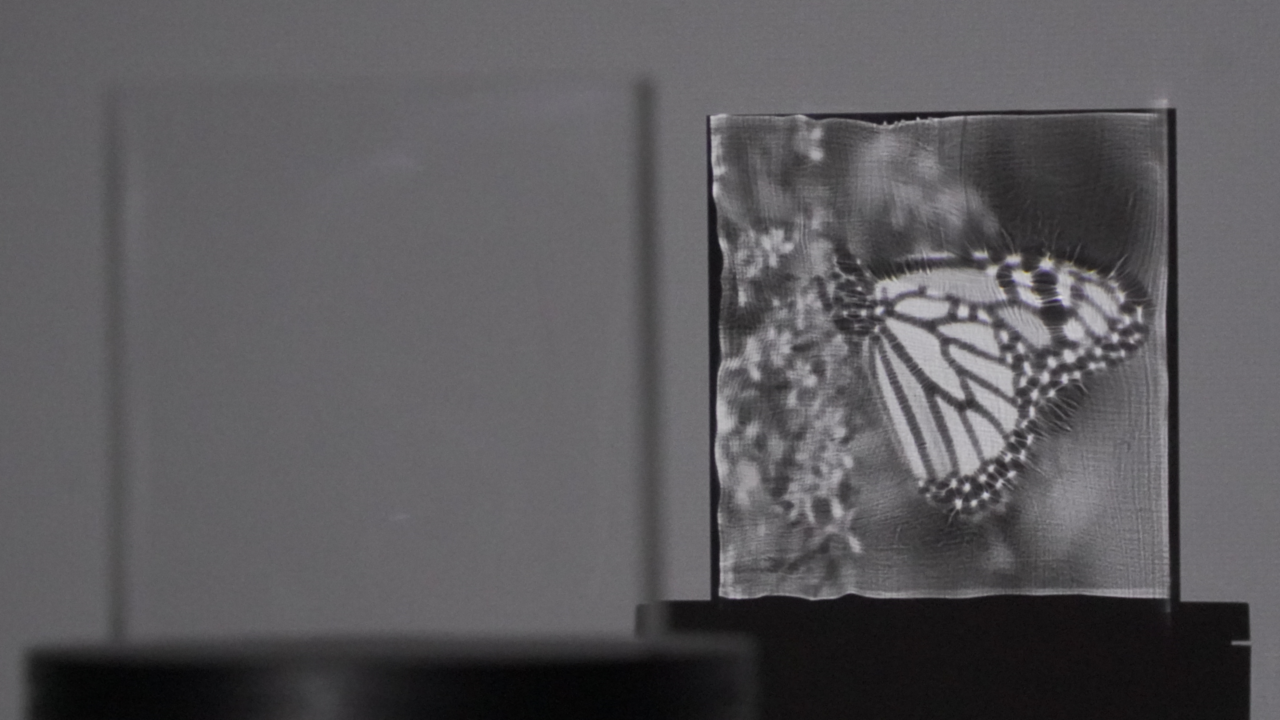}
\end{minipage}
\addhspace
\begin{minipage}{0.225\linewidth}
\includegraphics[width=\linewidth]{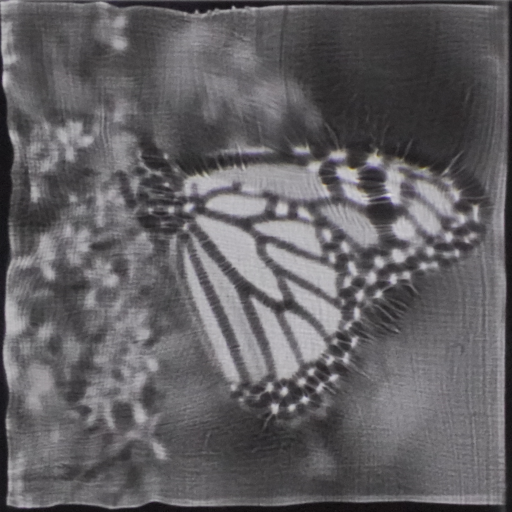}
\end{minipage}
\addhspace
\begin{minipage}{0.225\linewidth}
\includegraphics[width=\linewidth]{Figs/Butterfly_Ours.png}
\end{minipage}\\[0.4em]
\begin{minipage}{0.1\linewidth}
\centering
\vfill
\small{Point Light}
\vfill
\end{minipage}
\addhspace
\begin{minipage}{0.4\linewidth}
\includegraphics[width=\linewidth]{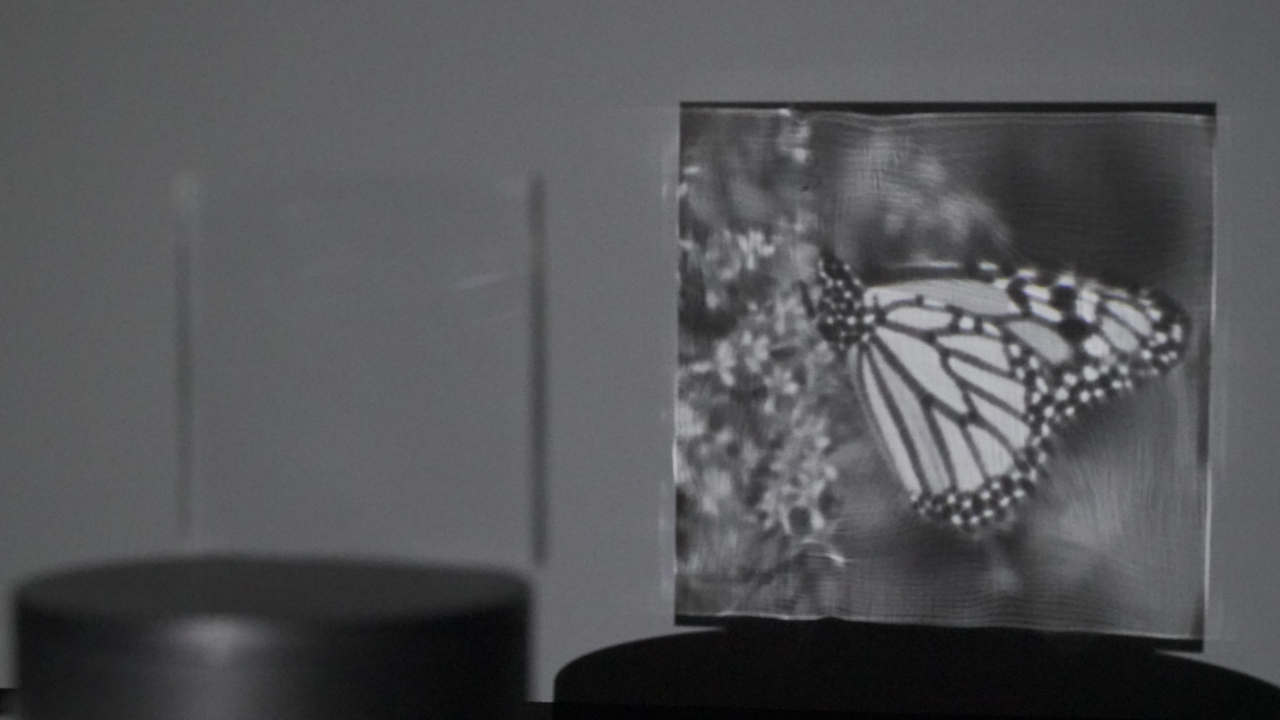}
\end{minipage}
\addhspace
\begin{minipage}{0.225\linewidth}
\includegraphics[width=\linewidth]{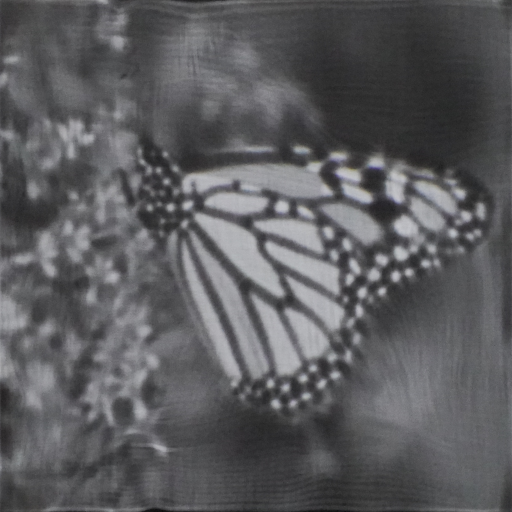}
\end{minipage}
\addhspace
\begin{minipage}{0.225\linewidth}
\includegraphics[width=\linewidth]{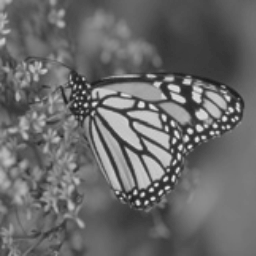}
\end{minipage}\\[0.4em]
\hspace{0.1\linewidth}
\addhspace
\begin{minipage}{0.4\linewidth}
\centering
\small{Physical Prototypes}
\end{minipage}
\addhspace
\begin{minipage}{0.225\linewidth}
\centering
\small{Rectified Results}
\end{minipage}
\addhspace
\begin{minipage}{0.225\linewidth}
\centering
\small{Rendered Results}
\end{minipage}
\caption{We fabricate physical prototypes for lens shapes optimized using our method for a parallel light source and a point light source. The photos in the column ``Rectified Results'' are produced by rectifying the caustic image within the leftmost photos, to provide a frontal view that facilitates comparison with the rendered results. Despite the deviation between our physical lights from ideal parallel and point lights and the imperfections arising from the fabrication process, the real images produced by the prototypes are close to the images generated from our rendering model.}
\label{fig:prototypes}
\end{figure*}

\begin{figure*}[t]
\centering
\begin{minipage}{0.464\linewidth}
\centerline{\includegraphics[width=1\linewidth]{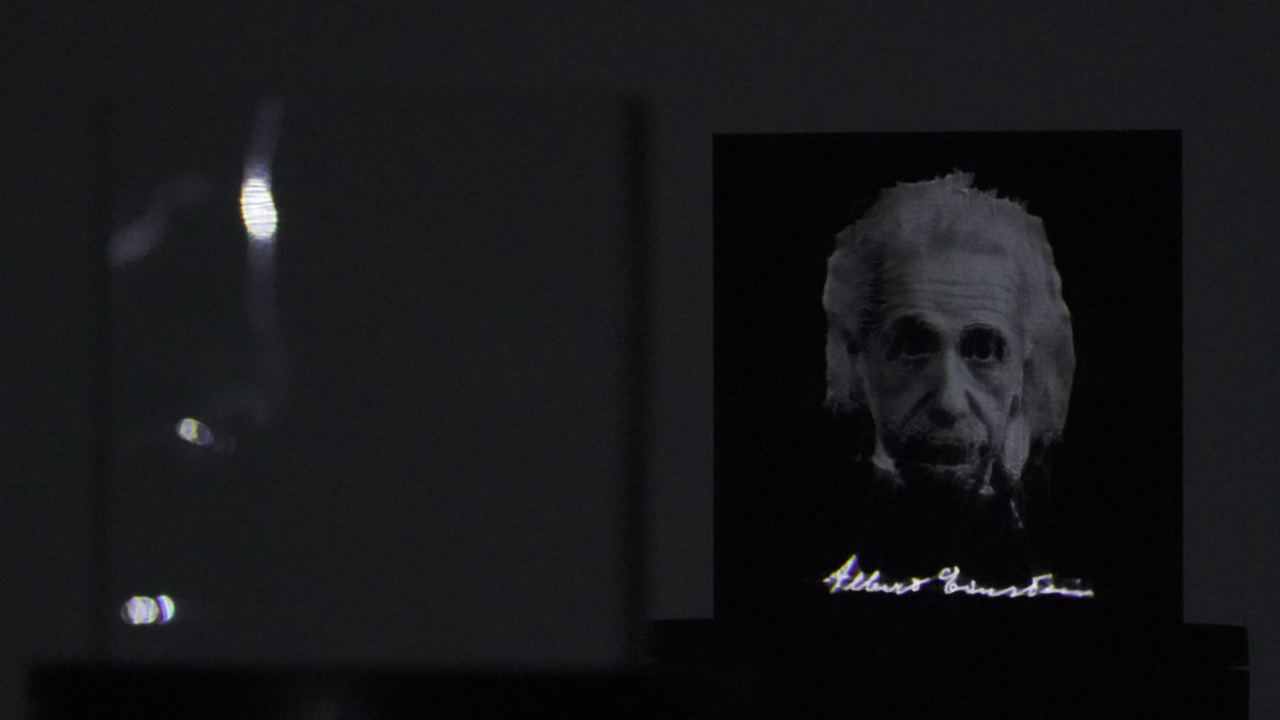}}
\end{minipage}
\hfill
\begin{minipage}{0.261\linewidth}
\centerline{\includegraphics[width=1\linewidth]{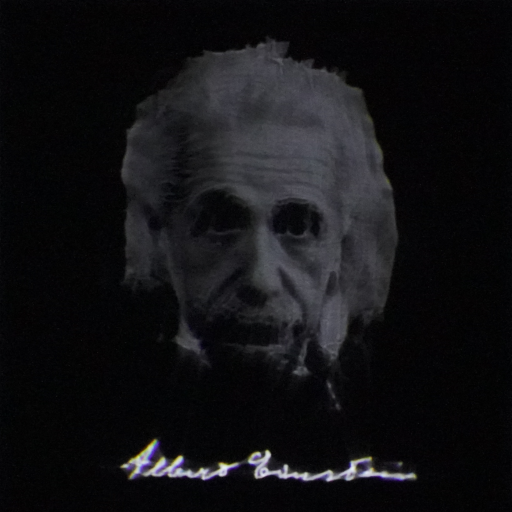}}
\end{minipage}
\hfill
\begin{minipage}{0.261\linewidth}
\centerline{\includegraphics[width=1\linewidth]{Figs/Einstein_Ours.png}}
\end{minipage}\\[0.2em]
\begin{minipage}{0.464\linewidth}
\vspace*{0.4em}
\centering
{\small Our Physical Prototype}
\end{minipage}
\hfill
\begin{minipage}{0.261\linewidth}
\vspace*{0.4em}
\centering
{\small  Rectified Result}
\end{minipage}
\hfill
\begin{minipage}{0.261\linewidth}
\vspace*{0.4em}
\centering
{\small  Our Rendered Result}
\end{minipage}\\[0.4em]
\begin{minipage}{0.464\linewidth}
\centerline{\includegraphics[width=1\linewidth]{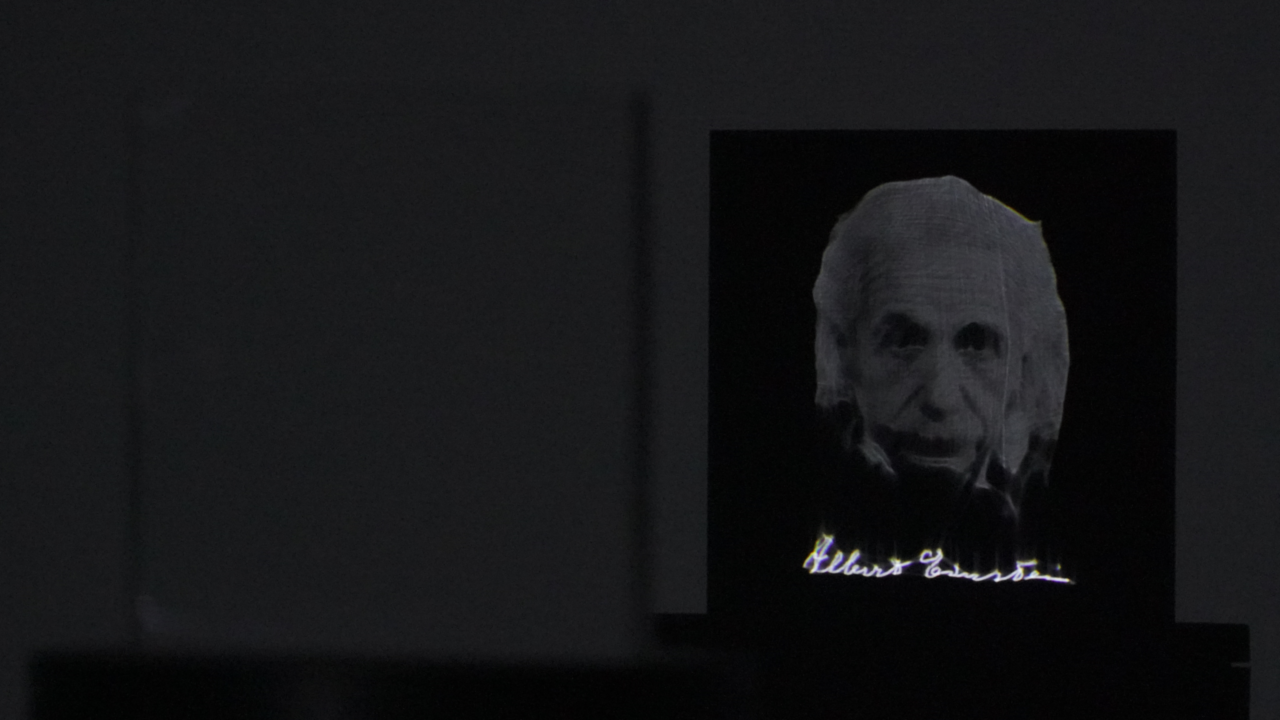}}
\end{minipage}
\hfill
\begin{minipage}{0.261\linewidth}
\centerline{\includegraphics[width=1\linewidth]{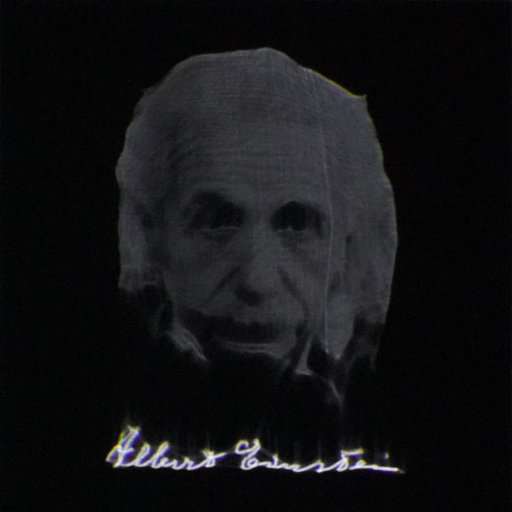}}
\end{minipage}
\hfill
\begin{minipage}{0.261\linewidth}
\centerline{\includegraphics[width=1\linewidth]{Figs/Einstein_gt.png}}
\end{minipage}\\[0.2em]
\begin{minipage}{0.464\linewidth}
\vspace*{0.4em}
\centering
{\small Physical Prototype Using \cite{schwartzburg2014high}}
\end{minipage}
\hfill
\begin{minipage}{0.261\linewidth}
\vspace*{0.4em}
\centering
{\small  Rectified Result}
\end{minipage}
\hfill
\begin{minipage}{0.261\linewidth}
\vspace*{0.4em}
\centering
{\small  Target Image}
\end{minipage}\\
\caption{Comparison of physical prototypes produced using our method and the method of~\cite{schwartzburg2014high}. The resulting mesh model of~\cite{schwartzburg2014high} was provided by the authors. Both results are physically realized using the same fabrication settings. 
The photos in the middle column are produced by rectifying the caustic image within the leftmost photos to a frontal view.
The real caustic image from the result of~\cite{schwartzburg2014high} exhibits more distortion compared to the target, and contains more fabrication artifacts. Einstein portrait \textcopyright~Magnum/IC photo.}
\label{fig:prototype_Einstein}
\end{figure*}

\begin{figure*}
\begin{minipage}{0.09\linewidth}
\centering{\small With\\Smoothness}
\end{minipage}
\hfill
\begin{minipage}{0.22\linewidth}
\centerline{\includegraphics[width=1\linewidth]{Figs/Einstein_Ours.png}}
\end{minipage}
\hfill
\begin{minipage}{0.22\linewidth}
\centerline{\includegraphics[width=1\linewidth]{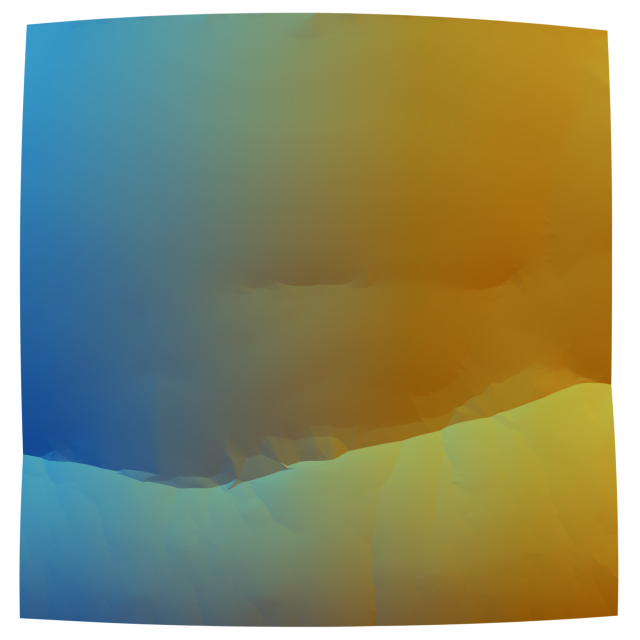}}
\end{minipage}
\hfill
\begin{minipage}{0.22\linewidth}
\centerline{\includegraphics[width=1\linewidth]{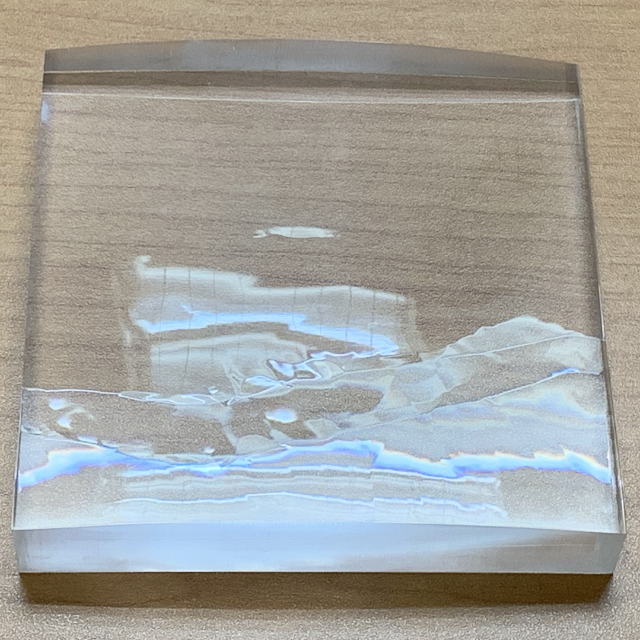}}
\end{minipage}
\hfill
\begin{minipage}{0.22\linewidth}
\centerline{\includegraphics[width=1\linewidth]{Figs/Einstein_pp_Correction.png}}
\end{minipage}
\vfill
\vspace*{0.4em}
\begin{minipage}{0.09\linewidth}
\centering{\small Without\\Smoothness}
\end{minipage}
\hfill
\begin{minipage}{0.22\linewidth}
\centerline{\includegraphics[width=1\linewidth]{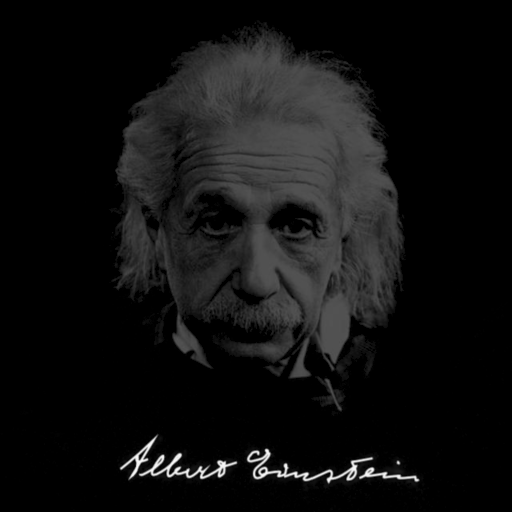}}
\end{minipage}
\hfill
\begin{minipage}{0.22\linewidth}
\centerline{\includegraphics[width=1\linewidth]{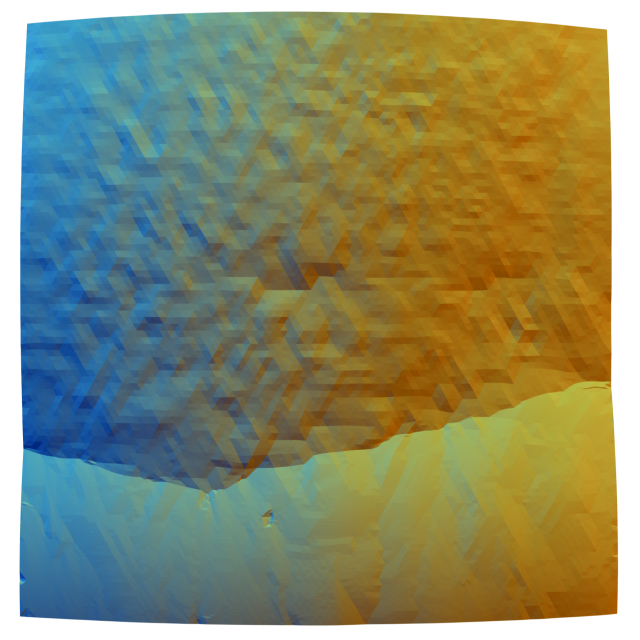}}
\end{minipage}
\hfill
\begin{minipage}{0.22\linewidth}
\centerline{\includegraphics[width=1\linewidth]{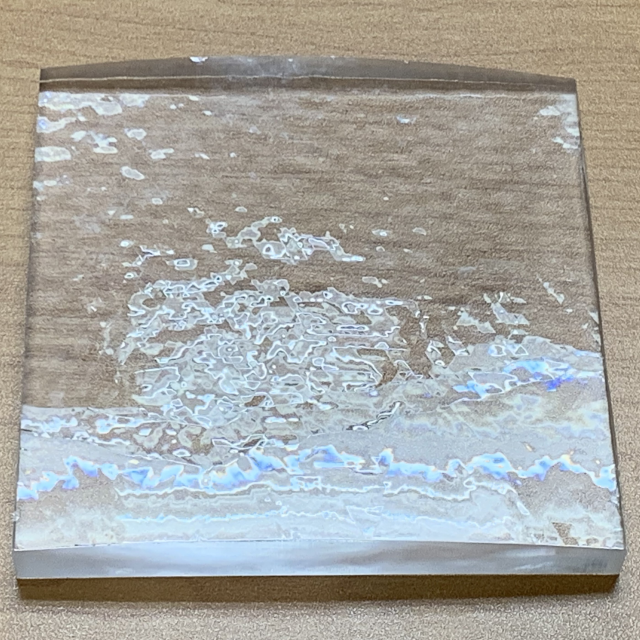}}
\end{minipage}
\hfill
\begin{minipage}{0.22\linewidth}
\centerline{\includegraphics[width=1\linewidth]{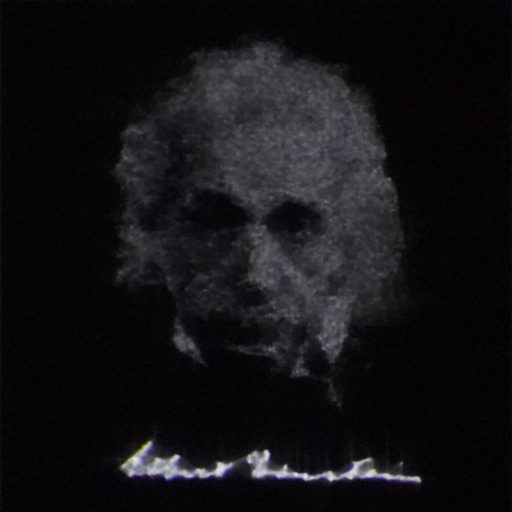}}
\end{minipage}
\vfill
\vspace*{0.4em}
\hspace{0.09\linewidth}
\hfill
\begin{minipage}{0.22\linewidth}
\centering{\small  Rendered Result}
\end{minipage}
\hfill
\begin{minipage}{0.22\linewidth}
\centering{\small Rendered Surface}
\end{minipage}
\hfill
\begin{minipage}{0.22\linewidth}
\centering{\small Fabricated Surface}
\end{minipage}
\hfill
\begin{minipage}{0.22\linewidth}
\centering{\small Rectified Physical Result}
\end{minipage}
\caption{We optimize and fabricate lens shapes using our method for the same target image, with and without the piecewise smoothness regularization. For the result without the smoothness term, there is a significant deviation between the caustic image rendered from the optimized surface and the real image produced by the physical prototype. Einstein portrait \textcopyright~Magnum/IC photo.}
\label{fig:prototype_smooth}
\end{figure*}

We showcase several physical prototypes in Fig.~\ref{fig:teaser}, Fig.~\ref{fig:prototypes}, Fig.~\ref{fig:prototype_Einstein} and Fig.~\ref{fig:prototype_smooth}. 
Our supplementary video also includes their demonstration. All models are fabricated in acrylic (IOR: 1.49) using a JDGR100 milling machine in 3-axis mode. After rough milling, the final pass is done using a 0.4mm ball-end mill with 0.06mm pitch at 23,000 RPM. Afterward, the pieces are polished manually using polishing wax and a cloth wheel. 
For all models with a parallel light source, we use a white LED torch placed approximately 10 meters away from the lens as the light source.
For the model with a point light course, we use a small white spherical LED (OSRAM OSLON SSL 150) with a diameter of about 3mm and positioned at 60cm from the lens.
It should be noted that such physical light sources inevitably deviate from the strictly uniform parallel light and the Lambertian point light that are assumed by our formulation, which can cause differences between the simulated rendering and the physical results.

Fig.~\ref{fig:teaser} and Fig.~\ref{fig:prototypes} show the rendering results of a few models created by our method, as well as caustic images produced by their physical prototypes.
To facilitate comparison with the rendered results, we have rectified the photos of the physical caustic images in Fig.~\ref{fig:prototypes} to provide a frontal view of the imaging area.
Despite the deviation between our physical light sources and their mathematical models, and the slight imperfections due to CNC tolerance and polishing, the physical prototypes still produce images with a strong resemblance to the target.

Fig.~\ref{fig:prototype_Einstein} compares physical results from our method and the method of~\cite{schwartzburg2014high} on the same target image. Both results are fabricated using the same settings. The caustic images in the real photos in the leftmost column have also been rectified to obtain a front view that facilitates comparison with the target image.  We can see that the result of~\cite{schwartzburg2014high} has notable distortions compared to the target image, especially in the silhouette region around the hairs. 
This is because in~\cite{schwartzburg2014high}, the lens shape is not directly optimized based on the rendering image; instead, they first compute a normal field matching the target image and then reconstruct a mesh from the normals. As their normal field is computed from OT alone, it is not guaranteed to be integrable; as a result, the normals of the reconstructed mesh may deviate from the OT normal field, causing distortions in the real image. In contrast, our optimization is directly driven by the rendering image and achieves better preservation of the target image. In addition, our physical prototype contains fewer fabrication artifacts in the resulting image, thanks to our piecewise smoothness constraints that facilitate fabrication.

Fig.~\ref{fig:prototype_smooth} further demonstrates the importance of our piecewise smoothness regularization. For the same Einstein target image, we compute and fabricate two lens shapes using our method with and without the smoothness term while keeping all other components the same. The optimized surface without the smoothness term is notably rougher and contains a large number of small concavities and convexities. Moreover, although both surfaces can produce images closely matching the target in simulated rendering, the real image from the physical prototype without the smoothness term deviates significantly from the target, while the one with the smoothness term still maintains strong resemblance to the target. This example shows that our piecewise smoothness regularization ensures the optimized surface can be physically fabricated with sufficient accuracy, helping to achieve a desirable final result.

While our method has demonstrated promising results in both simulated renderings and physical prototypes, there are still discrepancies between the two due to various factors in the fabrication process. 
First, our assumption of ideal light sources (strictly uniform parallel light or Lambertian point light) deviates from the physical light sources used in our experiments, which can lead to differences in the observed caustic patterns. 
Second, the finite precision of the CNC machine and the use of a spherical milling tool introduce approximations in reproducing the optimized surface geometry. In particular, the milling tool cannot perfectly reproduce sharp concave features, resulting in unintended light leakage between adjacent bright areas in the real caustic image (e.g., the SIGGRAPH Logo example in Fig.~\ref{fig:prototypes}).
Third, the manual polishing process can cause additional distortions due to non-uniform material removal and over-polishing. For example, in the second row of Fig.~\ref{fig:prototypes}, over-polishing has altered the surface normals at the boundary of the back surface and caused the refracted light rays to deviate from their intended direction and move toward the interior of the image, resulting in notable black areas along the image boundary. It is important to note that such fabrication imperfections are not accounted for in our optimization pipeline, which assumes an ideal manufacturing process. Incorporating these real-world constraints and limitations into the computational design framework is an important direction for future research, as will be discussed in Section~\ref{sec:conclusion}.

\section{Conclusion and Discussion}
\label{sec:conclusion}
We presented an end-to-end optical surface optimization algorithm directly driven by the difference between the rendering result and the target image. To solve this challenging problem, we proposed a hybrid strategy combining optimal transport for global structure alignment and rendering guided optimization for local refinement. To ensure the quality of the final physical result, we also introduce a piecewise smoothness regularization to facilitate CNC milling and polishing of the designed surface. Extensive experiments verified the effectiveness of our method.

Our method still has some limitations that we plan to address in future work. First, although our piecewise smoothness regularization helps to improve the consistency between the simulated rendering and physical results, imperfections can still arise during fabrication. In the future, we plan to extend our approach to fully consider the impact of fabrication in the optimization, i.e., to formulate an optimization based on fabrication results instead of simulated rendering. One potential approach is to include the milling and polishing processes into our differentiable pipeline, and jointly optimize the lens shape and the CNC tool path to reduce artifacts in physical prototypes. 
In addition, the mesh connectivity is fixed in the current optimization pipeline, which limits the expressive ability of the surface to some extent. Therefore, dynamic update of the mesh connectivity is another research direction worth exploring.

\begin{acks}
This research was supported by the National Natural Science Foundation of China (No.62441224, No.62272433) and the USTC Fellowship (No.S19582024). The numerical calculations in this paper have been done on the supercomputing system in the Supercomputing Center of University of Science and Technology of China.
\end{acks}

% \begin{acks}
% Thanks Professor Huichun Ye in USTC for fabrication.
% \end{acks}

% \balance
% \nocite{*}
\bibliographystyle{ACM-Reference-Format}
\bibliography{references}

\appendix

\newcommand{\scinum}[2]{{#1}\times 10^{#2}}

\section{Derivation of Refraction Direction}
\label{appx:refraction}

Given the incident ray direction $\incomingdir$ and 
the surface normal $\normal$ that are both unit vectors, we want 
to compute a refracted ray direction vector $\overline{\outgoingdir}$ such that the vector
$$
\outgoingorthodir = \overline{\outgoingdir} - \normal
$$
is orthogonal to $\normal$ (see inset figure).

\begin{wrapfigure}{r}[1em]{0.18\columnwidth}
	\vspace*{-1.6em}
	\hspace*{-1.2em}
	\includegraphics[width=0.18\columnwidth]{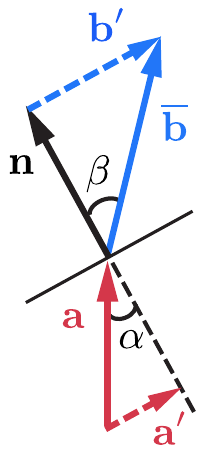}
\end{wrapfigure}
We first note that the vector
$$
\incomingorthodir = \incomingdir -  (\incomingdir \cdot \normal) \normal
$$
is parallel to $\outgoingorthodir$, and the ratio between their norms satisfies 
\[
    \frac{\|\outgoingorthodir\|}{\|\incomingorthodir\|}
    = \frac{\tan \outgoingangle}{\sin \incomingangle}
\]
since $\incomingdir$ and $\normal$ are both unit vectors. Using Snell's law, we have $\sin \outgoingangle = \riratio \sin \incomingangle$. Therefore,
$$
\outgoingorthodir = \frac{\tan \outgoingangle}{\sin \incomingangle} \incomingorthodir  =  \frac{\sin \outgoingangle}{\sin \incomingangle \cos \outgoingangle} \incomingorthodir =  \frac{\riratio}{\cos \outgoingangle} \incomingorthodir.
$$
Thus, we have 
\begin{equation*}
\overline{\outgoingdir} = \normal + \outgoingorthodir = \normal +  \frac{\riratio}{\cos \outgoingangle} \incomingorthodir = \frac{\normal \cos \outgoingangle   + \riratio \incomingorthodir}{\cos \outgoingangle}.
\end{equation*}
We can ignore the scaling factor in the denominator and simply use
\begin{equation}
    \outgoingdir = \normal \cos \outgoingangle   + \riratio \incomingorthodir
     = \normal \cos \outgoingangle + \riratio (\incomingdir -  (\incomingdir \cdot \normal) \normal )
     \label{eq:bVector}
\end{equation}
to represent the refracted ray direction. Moreover,
$\cos \outgoingangle$ can be computed from  Snell's law as:
\begin{equation}
\begin{aligned}
    \cos \outgoingangle &= \sqrt{1 - \sin^2 \outgoingangle} = \sqrt{1 - \riratio^2 \sin^2 \incomingangle} \\
    &= \sqrt{1 + \riratio^2(\cos^2 \incomingangle - 1)} = \sqrt{1 + \riratio^2((\normal \cdot \incomingdir)^2 - 1)}.
\end{aligned}
\label{eq:cosbeta}
\end{equation}
Substituting Eq.~\eqref{eq:cosbeta} into Eq.~\eqref{eq:bVector}, we have
\begin{equation}
\label{eq:Outgoing}
    \outgoingdir = \normal \sqrt{1 + \riratio^2((\normal \cdot \incomingdir)^2 - 1)} + \riratio (\incomingdir - (\incomingdir \cdot \normal) \normal).
\end{equation}

\section{Proof of No Re-intersection between Refracted Light Rays and the Lens Surface}
\label{appx:NoIntersection}

In the following, we show that the target function term $\singletirloss$ in Eq.~\eqref{eq:tir} ensures that an outgoing refracted ray from the back surface of the lens will not intersect with the surface again.

\begin{wrapfigure}{r}[1em]{0.45\columnwidth}
	%\vspace*{-1.3em}
	\hspace*{-1.6em}
	\includegraphics[width=0.45\columnwidth]{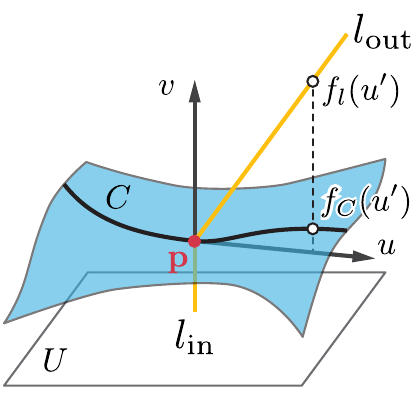}
\end{wrapfigure}
Recall that the back surface of the lens is represented as a height field over a planar domain $U$ that is identified with the front surface, and the light rays hitting the back surface are orthogonal to $U$. For any point $\mathbf{p}$ on the back surface, let $l_{\text{in}}$ and $l_{\text{out}}$ be the incoming and outgoing light rays, respectively. If $l_{\text{out}}$ is parallel to $l_{\text{in}}$, then $l_{\text{out}}$ is also orthogonal to $U$, and $\mathbf{p}$ will be the only intersection point between $l_{\text{out}}$ and the back surface because the back surface is a height field over $U$.

If $l_{\text{out}}$ is not parallel to $l_{\text{in}}$, then any intersection point between $l_{\text{out}}$ and the back surface must lie on the intersection curve $C$ between the back surface and the plane $P$ spanned by $l_{\text{in}}$ and $l_{\text{out}}$. In the following, we will show that $\mathbf{p}$ is the only intersection point between $l_{\text{out}}$ and $C$, and so it is also the only intersection point between $l_{\text{out}}$ and the back surface. To do so, we first construct 2D coordinate system $(u, v)$ within the plane $P$ such that (see inset figure):
\begin{itemize}
    \item the origin of the coordinate system is at $\mathbf{p}$;
    \item the $u$-axis is parallel to the front surface, while the $v$-axis is parallel to $l_{\text{in}}$ and points to the receptive plane;
    \item the ray $l_{\text{out}}$ is represented with an equation 
    \begin{equation}
    v = f_l(u) = K u, \qquad u \geq 0,
    \label{eq:OutRay2D}
    \end{equation}
    where $K > 0$ is the slope of $l_{\text{out}}$.
\end{itemize}
Within this coordinate system, the curve $C$ can be represented via $$v = f_C(u)$$ where $f_C(0) = 0$, and we only need to show that there is no intersection between $l_{\text{out}}$ and $C$ on the interval $u \in (0, +\infty)$.

To this end, we note that due to the term $\singletirloss$ in Eq.~\eqref{eq:tir}, the unit normal vector $\mathbf{n}$ at any point on the back surface must satisfy
\begin{equation}
    (\mathbf{n} \cdot \mathbf{a})^2 > 1 - \eta^{-2},
    \label{eq:NormalDirConstraint}
\end{equation}
where $\mathbf{a}$ is the unit vector for the incoming light direction and coincides with the +$v$ axis direction in the 2D coordinate system mentioned above.
As the back surface is a height-field surface, we must have $\mathbf{n} \cdot \mathbf{a} > 0$. Therefore, Eq.~\eqref{eq:NormalDirConstraint} implies that 
\[
    \mathbf{n} \cdot \mathbf{a} > \sqrt{1 - \eta^{-2}},
\]
meaning that the angle between the back surface normal and the vector $\mathbf{a}$ must be smaller than a bound 
\begin{equation}
    \theta_{\max} = \arccos(\sqrt{1 - \eta^{-2}}) = \arcsin(\eta^{-1})
    \label{eq:theta_max}
\end{equation}
at any point on the back surface. As a result, at any point on the back surface, the angle $\theta'$ from any line in its tangent plane to the front surface plane
must satisfy
\begin{equation}
    0 \leq \theta' <  \theta_{\max}.
    \label{eq:TangentLineAngleBound}
\end{equation}
Moreover, for the local height function $f_C(u)$ of the curve $C$, the derivative $f'_C(u)$ satifies
\[
    |f'_C(u)| = \tan(\theta_u)
\]
where $\theta_u$ is the angle between the tangent line of $C$ at $u$ and the front surface plane. Then, Eq.~\eqref{eq:TangentLineAngleBound} means that
\[
    |f'_C(u)| < \tan(\theta_{\max}).
\]
As a result, for any $u' > 0$, we have
\begin{equation}
    f_C(u') = f_C(0) + \int_0^{u'} f'_C(u) du \leq \int_0^{u'} |f'_C(u)| du < u' \cdot \tan(\theta_{\max}).
    \label{eq:fCBound}
\end{equation}

Meanwhile, at any point on the back surface, the angle between the refracted ray direction and the back surface normal must be no larger than $\pi/2$, since the term $\singletirloss$ ensures that there is no total internal reflection. Therefore, the angle $\theta_{\text{out}}$ between the refracted ray and the front surface plane must satisfy
\begin{equation}
    \theta_{\text{out}} > \theta_{\max}.
\end{equation}
As a result, the slope $K$ of ray $l_{\text{out}}$ in Eq.~\eqref{eq:OutRay2D} must satisfy
\[
    K > \tan(\theta_{\max}),
\]
so we have 
\begin{equation}
    f_l(u') = K u' >  u' \cdot \tan(\theta_{\max})
    \label{eq:fOutBound}
\end{equation}
for any $u' > 0$. Then, using Eqs.~\eqref{eq:fCBound} and \eqref{eq:fOutBound}, we have
\[
    f_l(u') > f_C(u'), \qquad \forall~u'>0.
\]
In other words, away from the point $\mathbf{p}$, the outgoing ray $l_{\text{out}}$ is always above the curve $C$. Therefore, $\mathbf{p}$ is the only intersection point between $l_{\text{out}}$ and the back surface.

\section{Evaluation of Energy $H(\mathbf{w})$}
\label{appx:H}
The main difficulty in evaluating  $H(\mathbf{w})$ comes from the integral
\begin{align*}
     & \int_{\mathbf{s} \in \powercell{i}{\mathbf{w}}} (\| \mathbf{x} - \mathbf{c}'_i \|^2 - w_i) \widetilde{\Phi}_\mathbf{s} d \mathbf{s}\\
    =~ & \int_{\mathbf{s} \in \powercell{i}{\mathbf{w}}} \| \mathbf{x} - \mathbf{c}'_i \|^2 \widetilde{\Phi}_\mathbf{s} d \mathbf{s}
    - w_i \int_{\mathbf{s} \in \powercell{i}{\mathbf{w}}} \widetilde{\Phi}_\mathbf{s}d\mathbf{s}.
\end{align*}
Since the other components in $H(\mathbf{w})$ are trivial to evaluate, in the following we focus on the computation of two integrals,
\begin{equation}
\label{eq2}
\int_{\mathbf{s} \in \powercell{i}{\mathbf{w}}} \widetilde{\Phi}_\mathbf{s}d\mathbf{s}
\end{equation}
and
\begin{equation}
\label{eq1}
\int_{\mathbf{s} \in \powercell{i}{\mathbf{w}}} \left\| \mathbf{x} - \mathbf{c}_i' \right\|^2 \widetilde{\Phi}_\mathbf{s}d\mathbf{s}.
\end{equation}

For Eq.~\eqref{eq2}, we compute the integral by accumulating the contribution from each pixel:
\begin{equation}
\label{sub_eq2}
    \begin{aligned}
        \int_{\mathbf{s} \in \powercell{i}{\mathbf{w}}}  \widetilde{\Phi}_\mathbf{s}d\mathbf{s} 
       & = \sum_{j = 1}^{n_p} \left(\int_{\mathbf{s} \in \powercell{i}{\mathbf{w}} \cap p^j} \frac{\widetilde{\Phi}^j}{A_p} d\mathbf{s}\right)\\
       & = \sum_{j = 1}^{n_p} \frac{\widetilde{\Phi}^j}{A_p} \left(\int_{\mathbf{s} \in \powercell{i}{\mathbf{w}} \cap p^j} d\mathbf{s}\right)\\
       & = \sum_{j = 1}^{n_p} \frac{\widetilde{\Phi}^j A(\powercell{i}{\mathbf{w}} \cap p^j)}{A_p},
    \end{aligned}
\end{equation}
where $A(\cdot)$ denotes the area. Note that since both $\powercell{i}{\mathbf{w}}$ and $p^j$ are both convex polygons, if their intersection is not empty then $\powercell{i}{\mathbf{w}} \cap p^j$ must be a convex polygon. It is easy to determine the intersection area $A(\powercell{i}{\mathbf{w}} \cap p^j)$ in this case.

Similarly, for Eq.~\eqref{eq1} we accumulate the contributions from all pixels:
\begin{equation}
\label{sub_eq1}
    \begin{aligned}
        & \int_{\mathbf{s} \in \powercell{i}{\mathbf{w}}} \left\| \mathbf{x} - \mathbf{c}_i' \right\|^2 \widetilde{\Phi}_\mathbf{s}d\mathbf{s} \\
        = & \sum_{j = 1}^{n_p} \left(\int_{\mathbf{s} \in \powercell{i}{\mathbf{w}} \cap p^j} \left\| \mathbf{x} - \mathbf{c}_i' \right\|^2 \frac{\widetilde{\Phi}^j}{A_p} d\mathbf{s}\right)\\
        = &\sum_{j = 1}^{n_p} \frac{\widetilde{\Phi}^j}{A_p} \int_{\mathbf{s} \in \powercell{i}{\mathbf{w}} \cap p^j} \left\| \mathbf{x} - \mathbf{c}_i' \right\|^2 d\mathbf{s}.
    \end{aligned}
\end{equation}
Since $\frac{\widetilde{\Phi}^j}{A_p}$ is a given constant for each $p^j$, we only need to consider the calculation of 
\begin{equation}
\int_{\mathbf{s} \in \powercell{i}{\mathbf{w}} \cap p^j} \left\| \mathbf{x} - \mathbf{c}_i' \right\|^2 d\mathbf{s}.
\label{eq:Integral}
\end{equation}

Recall the following fact from the Green's Theorem: if both $P$ and $Q$ have have continuous partial derivatives on the domain $D$ and its boundary $L$, then 
\begin{equation}
	\iint_D \left(\frac{\partial Q}{\partial x} - \frac{\partial P}{\partial y}\right) = \oint_{L^+} Pdx + Qdy.
 \label{eq:GreenTheorem}
\end{equation}
We would like to apply this relation to convert the integral in Eq.~\eqref{eq:Integral} into a line integral. To this end, we need to find two functions $P$ and $Q$, such that 
$$
\frac{\partial Q}{\partial x} - \frac{\partial P}{\partial y} = \left\| \mathbf{x} - \mathbf{c}_i' \right\|^2.
$$
A symmetric solution is $Q = x(y - y_0)^2$ and $P = -y(x - x_0)^2$ where $(x_0, y_0) = \mathbf{c}_i'$. 
Then we can apply Eq.~\eqref{eq:GreenTheorem} to each domain $\powercell{i}{\mathbf{w}} \cap p^j$. Note that $\powercell{i}{\mathbf{w}} \cap p^j$ is a convex polygon if it is not empty. Therefore, we only need to consider the right-hand-side of \eqref{eq:GreenTheorem} on each segment of its boundary.
Suppose $(x_1, y_1)$ and $(x_2, y_2)$ are two adjacent vertices of this polygon in counterclockwise order, then we have
\begin{equation}
\label{Qdy}
\begin{aligned}
&\int_{y_1}^{y_2}Q dy\\ 
=~ &\int_{y_1}^{y_2}x(y - y_0)^2 dy\\
=~ &\int_{y_1}^{y_2}(ky + b)(y - y_0)^2 dy\\
=~ &\int_{y_1}^{y_2}ky^3 + (b-2ky_0)y^2 + (ky_0^2-2by_0)y + by_0^2\ dy\\
=~ &\left.\left(\frac{k}{4}y^4 + \frac{b - 2ky_0}{3}y^3 + \frac{ky_0^2-2by_0}{2}y^2 + by_0^2 y\right) \right|_{y_1}^{y_2}\\
=~ &\frac{x_2 - x_1}{4}(y_1 + y_2)(y_1^2 + y_2^2) \\
& +\frac{x_1y_2 - x_2y_1-2y_0(x_2 - x_1)}{3}(y_1^2 + y_2^2 + y_1y_2) \\
& + \frac{y_0^2(x_2 - x_1) - 2 y_0 (x_1y_2 - x_2 y_1)}{2}(y_1 + y_2) \\
& + y_0^2(x_1y_2 - x_2 y_1).
\end{aligned} 
\end{equation}
Here we make use of the linear relation $x = ky+b$, where $k=\frac{x_2 - x_1}{y_2 - y_1}$ and $b = \frac{x_1 y_2 - x_2 y_1}{y_2 - y_1}$.

Similarly, we can deduce:
\begin{equation}
\label{Pdx}
\begin{aligned}
&\int_{x_1}^{x_2}P dx \\
=~ &\int_{x_1}^{x_2} -y(x - x_0)^2 dx\\
=~ &\int_{x_1}^{x_2}(mx + c)(x - x_0)^2 dx\\
=~ &\int_{x_1}^{x_2}mx^3 + (c-2mx_0)x^2 + (mx_0^2-2cx_0)x\  + c x_0^2\ dx\\
=~ &\left.\left(\frac{m}{4}x^4 + \frac{c - 2mx_0}{3}x^3 + \frac{mx_0^2-2cx_0}{2}x^2 + c x_0^2 x\right)\right|_{x_1}^{x_2}\\
=~ &\frac{y_1 - y_2}{4}(x_1 + x_2)(x_1^2 + x_2^2) \\
& +\frac{x_1 y_2 - x_2 y_1 + 2 x_0 (y_2 - y_1)}{3}(x_1^2 + x_2^2 + x_1 x_2)\\
& +\frac{x_0^2(y_1 - y_2) - 2x_0(x_1 y_2 - x_2 y_1)}{2}(x_1 + x_2) \\
& +x_0^2 (x_1 y_2 - x_2 y_1), 
\end{aligned}
\end{equation}
where we make use of the linear relation $-y=mx+c$, with $m=-\frac{y_2 - y_1}{x_2 - x_1} = -\frac{1}{k}$ and $c = \frac{x_1 y_2 - x_2 y_1}{x_2 - x_1}=\frac{b}{k}$.

To summarize, for computing the integral~\eqref{eq:Integral}, we apply Eqs.~\eqref{Qdy} and \eqref{Pdx} to each boundary edge of $\powercell{i}{\mathbf{w}} \cap p^j$ in the counter-clockwise direction and sum their values.
In addition, we only consider pixels inside the bounding box of $\powercell{i}{\mathbf{w}}$, to avoid unnecessary intersection.

\section{Extension to Reflection}
\label{appx:reflection}
For reflection surfaces, we can also set the starting surface to a plane with a normal $\mathbf{n} = (0, 0, 1)$.
When setting the light source, we must also ensure the light does not strike the reflective surface straight on, as this would cause the reflected light to align with the incoming light. In those cases, if a receptive plane is placed in the path of the reflected light, it could block the incoming light.
Typically, for parallel light we can set an inclination angle between the incoming light direction and the initial surface normal, and place the receptive plane on the other side of the normal to avoid this issue. 

At a surface point with a surface normal $\mathbf{n}$ and an incident light direction $\mathbf{a}$, it is easy to derive the reflected light direction as
\[
\mathbf{b} = \mathbf{a} - 2 (\mathbf{a} \cdot \mathbf{n}) \mathbf{n}.
\]

\section{Further Details of Implementation}
\label{appx:details}

In this section, we present more details of our implementation. 

For all examples, we first downsample the target image to a resolution of $32^2$ ($50 \times 25$ for the Calligraphy image) and use the coarsest mesh as the initial stage. The mesh resolution is independent of the image resolution, which means that we can freely choose the mesh resolution based on the tradeoff between optimization time and expressive power. In our examples, we set the mesh resolution to 1, 1.25, or 2.5 times the image resolution depending on our needs. For the butterfly example, we used a mesh with a resolution of 2.5 times the image resolution, which resulted in a very small MAE. 

We use 4 threads of an Intel Xeon E5-2690 to compute the power diagram using CGAL, and an NVIDIA GeForce RTX 3090 for all the optimizations. At the highest resolution, it takes approximately two hours to perform OT, five minutes for correspondence guided update, and an hour for rendering guided optimization.

In $\edgesmoothloss$, the parameter $\nu$ for the Welsch function controls its sensitivity to outliers. The smaller it is, the closer the Welsch function is to the $\ell_0$-norm, and the less influential the outliers are to the optimization. For the sharp edges to emerge, it is necessary to set $\nu$ to a sufficiently small value. On the other hand, setting a small $\nu$ from the beginning could prevent the optimization from improving edge smoothness, as the errors at most edges could be treated as outliers and have little influence on the optimization. To facilitate optimization, we set a relatively large $\nu$ at the beginning so that all edges can affect the optimization, then gradually decrease it to a small value to allow sharp edges to emerge. Specifically, we first set the maximum and minimum values 
$\alpha_{\max}, \alpha_{\min}$ of the ratio between $\nu$ and the typical scale $D_e$ of the residual $\sqrt{\edgeerrfunc(e_{ij})}$ on an edge. 
We set $\nu = \alpha_{\max} \cdot D_e$ at the beginning of the optimization. Afterward, when the mesh is subdivided, we decrease $\nu$ by a fixed ratio
$\gamma = (\displaystyle\frac{\alpha_{\min}}{\alpha_{\max}})^{\frac{1}{k-1}}$ where $k$ is the number of mesh resolutions,
such that $\nu$ is set to $\alpha_{\min} \cdot D_e$ for the highest resolution.

In our rendering guided optimization, the weights in the optimization target function are set to
$$
(\targetweightone_1, \targetweightone_2, \targetweightone_3, \targetweightone_4, \targetweightone_5) = (\scinum{1}{2}, \scinum{1}{3}, \scinum{1}{-3}, \scinum{2}{1}, \scinum{1}{-8}).
$$ 
For the correspondence guided update, we set 
$$(\targetweighttwo_1, \targetweighttwo_2, \targetweighttwo_3, \targetweighttwo_4) = (\scinum{1}{0}, \scinum{1}{1}, \scinum{2}{1}, \scinum{1}{-14}).$$ 
For the piecewise smoothness term $\fullsmoothloss$, we set 
$$
(\smoothweight_1, \smoothweight_2) = (\scinum{3}{0}, \scinum{2}{-1}).
$$

\end{document}